\documentclass[twocolumn]{aastex631}

\usepackage{lineno}
\usepackage{multirow}
\usepackage{threeparttable}
\usepackage{xspace}

\newcommand{\feh}{[Fe/H]}

\newcommand{\gruDM}{18.59}
\newcommand{\gruDMerr}{0.08}
\newcommand{\gruDKPC}{52.3}
\newcommand{\gruDKPCerr}{1.9}
\newcommand{\gruRA}{331.0237}
\newcommand{\gruDEC}{$-$46.4509}
\newcommand{\gruELLIP}{0.25}
\newcommand{\gruELLIPerr}{0.07}
\newcommand{\gruRH}{6.8}
\newcommand{\gruRHerr}{0.5}
\newcommand{\gruRHPHYS}{103}
\newcommand{\gruRHPHYSerr}{9}
\newcommand{\gruTHETA}{186}
\newcommand{\gruMV}{$-$4.07}
\newcommand{\gruMVerr}{0.50}
\newcommand{\gruPMRA}{0.384}
\newcommand{\gruPMRAerr}{0.033}
\newcommand{\gruPMDEC}{$-$1.484}
\newcommand{\gruPMDECerr}{0.040}

\newcommand{\horoDM}{19.30}
\newcommand{\horoDMerr}{0.17}
\newcommand{\horoDKPC}{72.4}

\newcommand{\horoRA}{49.1344}
\newcommand{\horoDEC}{$-$50.0070}
\newcommand{\horoELLIP}{0.32}
\newcommand{\horoRH}{2.1}
\newcommand{\horoRHerr}{0.2}
\newcommand{\horoRHPHYS}{44}
\newcommand{\horoTHETA}{144}
\newcommand{\horoMV}{$-$2.10}
\newcommand{\horoMVerr}{0.44}
\newcommand{\horoPMRA}{0.967}
\newcommand{\horoPMRAerr}{0.173}
\newcommand{\horoPMDEC}{$-$0.771}
\newcommand{\horoPMDECerr}{0.230}

\submitjournal{ApJ}

\shorttitle{Deep imaging of Gru~II and Hor~II}
\shortauthors{Prabhu et al.}

\newcommand{\UA}{\affiliation{Steward Observatory, University of Arizona, 933 North Cherry Avenue, Tucson, AZ 85721-0065, USA}}

\newcommand{\Dart}{\affiliation{Department of Physics and Astronomy, Dartmouth College, 6127 Wilder Laboratory, Hanover, NH 03755, USA}}

\begin{document}

\title{Deep Imaging of Grus~II and Horologium~II: Structure and Extent of Two Ultra-Faint Milky Way Satellites\footnote{This paper includes data gathered with the 6.5 m Magellan Telescopes at Las Campanas Observatory, Chile.}}

\correspondingauthor{Deepthi S. Prabhu}

\author[0000-0002-8217-5626]{Deepthi S. Prabhu}
\UA
\email{dprabhu@arizona.edu}

\author[0000-0003-4102-380X]{David J. Sand}
\UA
%email{dsand@arizona.edu}

\author[0000-0002-7155-679X]{Anirudh Chiti}
\affil{Kavli Institute for Particle Astrophysics \& Cosmology, Stanford University, Stanford, CA 94305, USA}
%\email{achiti@stanford.edu}

\author[0000-0001-9649-4815]{Bur\c{c}in Mutlu-Pakdil}
%\email{Burcin.Mutlu-Pakdil@dartmouth.edu}
\Dart

\author[0009-0007-9488-7050]{Sasha N. Campana}
%\email{sncampana@aol.com}
\Dart

\author[0000-0002-3936-9628]{J.~L.~Carlin}
\affiliation{Rubin Observatory/AURA, 950 North Cherry Avenue, Tucson, AZ, 85719, USA}
%\email{jeffreylcarlin@gmail.com}

\author[0000-0002-4863-8842]{A. P. Ji}
\affil{Department of Astronomy \& Astrophysics, University of Chicago, Chicago, IL 60637, USA}
\affil{Kavli Institute for Cosmological Physics, University of Chicago, Chicago, IL 60637, USA}
\affil{NSF-Simons AI Institute for the Sky (SkAI),172 E. Chestnut St., Chicago, IL 60611, USA}
%\email{alexji@uchicago.edu}

\author[0000-0002-4350-7632]{Jaclyn Jensen}
%\email{Jaclyn.R.Jensen@dartmouth.edu}
\Dart

\author[0000-0002-9144-7726]{C.~E.~Mart\'inez-V\'azquez}
%\email{clara.martinez@noirlab.edu}
\affiliation{NSF NOIRLab, 670 N. A'ohoku Place, Hilo, Hawai'i, 96720, USA}

\author[0000-0002-5177-727X]{Dennis Zaritsky}
\affiliation{Steward Observatory, University of Arizona, 933 North Cherry Avenue, Tucson, AZ 85721-0065, USA}
%\email{dfz@arizona.edu}

\author[0000-0002-6021-8760]{A.~B.~Pace}
%\email{pvpace1@gmail.com}
\affiliation{Department of Astronomy, University of Virginia, 530 McCormick Road, Charlottesville, VA 22904, USA}

\author[0000-0001-5805-5766]{A.~H.~Riley}
\affiliation{ Lund Observatory, Division of Astrophysics, Department of Physics, Lund University, SE-221 00 Lund, Sweden}
%\email{alexander.riley@fysik.lu.se}

\author[0000-0002-1763-4128]{D.~Crnojevi\'c}
\affiliation{Department of Physics \& Astronomy, University of Tampa, 401 West Kennedy Boulevard, Tampa, FL 33606, USA}
%\email{dcrnojevic@ut.edu}

\author[0000-0002-9269-8287]{G.~Limberg}
 \affiliation{Kavli Institute for Cosmological Physics, University of Chicago, Chicago, IL 60637, USA}
 %\email{limberg@uchicago.edu}

\author[0000-0001-5368-3632]{Laura Congreve Hunter}
\affil{Department of Physics and Astronomy, Dartmouth College, Hanover, NH 03755, USA}
%\email{Laura.C.Hunter@dartmouth.edu}

\author[0000-0002-0956-7949]{Kristine Spekkens}
\affiliation{Department of Physics, Engineering Physics and Astronomy, Queen’s University, Kingston, ON K7L 3N6, Canada}
%\email{kristine.spekkens@gmail.com}

\author[0000-0002-5434-4904]{Michael G. Jones}
\affiliation{Steward Observatory, University of Arizona, 933 North Cherry Avenue, Tucson, AZ 85721-0065, USA}
%\email{mgjones@ipac.caltech.edu}

\author[0000-0001-9775-9029]{Amandine Doliva-Dolinsky}
\affiliation{Department of Physics, University of Surrey, Guildford GU2 7XH, UK}
%\email{a.doliva-dolinsky@surrey.ac.uk}

\author[0000-0001-8354-7279]{Paul Bennet}
\affiliation{Space Telescope Science Institute, 3700 San Martin Drive, Baltimore, MD 21218, USA}
%\email{pbennet@stsci.edu}

\author[0000-0003-4479-1265]{V. M. Placco}
\affil{NSF NOIRLab, Tucson, AZ 85719, USA}
%\email{vinicius.placco@noirlab.edu}

\author[0009-0005-9002-4800]{Quinn O. Casey}
%\email{quinn.o.casey.gr@dartmouth.edu}
\Dart

\author[0000-0003-4394-7491]{Guinevere Herron}
\Dart
%\email{Guinevere.Herron.GR@dartmouth.edu}

\author[0000-0003-1697-7062]{W.~Cerny}
\affiliation{Department of Astronomy, Yale University, New Haven, CT 06520, USA}
%\email{william.cerny@yale.edu}

\author[0000-0002-3204-1742]{Nitya Kallivayalil}
\affiliation{Department of Astronomy, University of Virginia, 530 McCormick Road, Charlottesville, VA 22904, USA}
%\email{njk3r@virginia.edu}

\author[0000-0003-1680-1884]{Y.~Choi}
\affiliation{NSF NOIRLab, 950 N. Cherry Ave., Tucson, AZ 85719, USA}
%\email{yumi.choi@noirlab.edu}

\author[0000-0003-1479-3059]{Guy S. Stringfellow}
\affiliation{University of Colorado Boulder, Boulder, CO 80309, USA}
%\email{Guy.Stringfellow@colorado.edu}

\author[0000-0002-1793-3689]{D.~L.~Nidever}
\affiliation{Department of Physics, Montana State University, P.O. Box 173840, Bozeman, MT 59717-3840}
\altaffiliation{NSF's National Optical-Infrared Astronomy Research Laboratory, 950 N. Cherry Ave., Tucson, AZ 85719, USA}
%\email{dnidever@montana.edu}

\author[0000-0002-3690-105X]{J.~A.~Carballo-Bello}
\affiliation{Instituto de Alta Investigaci\'on, Universidad de Tarapac\'a, Casilla 7D, Arica, Chile}
%\email{jcarballo@academicos.uta.cl}

\author[0000-0001-9438-5228]{M.~Navabi}
\affiliation{Department of Physics, University of Surrey, Guildford GU2 7XH, UK}
%\email{m.navabi@surrey.ac.uk}

\author[0000-0002-8093-7471]{P.~Massana}
\affiliation{NSF NOIRLab, Casilla 603, La Serena, Chile}
%\email{pol.massana@noirlab.edu}

\author[0000-0003-0105-9576]{G.~E.~Medina}
\affiliation{David A. Dunlap Department of Astronomy \& Astrophysics, University of Toronto, 50 St George Street, Toronto ON M5S 3H4, Canada}
\affiliation{Dunlap Institute for Astronomy \& Astrophysics, University of Toronto, 50 St George Street, Toronto, ON M5S 3H4, Canada}
%email{gustavo.medina@utoronto.ca}

\author[0000-0002-8282-469X]{N.~E.~D.~No\"el}
\affiliation{Department of Physics, University of Surrey, Guildford GU2 7XH, UK}
%\email{n.noel@surrey.ac.uk}

\collaboration{32}{(MAGIC \& DELVE Collaborations)}

\begin{abstract}

We present deep, wide-field Magellan/Megacam imaging of the  ultra-faint Milky Way (MW) satellites Grus~II (Gru~II) and Horologium~II (Hor~II), with the aim of deriving improved constraints on their distances, luminosities, and structural parameters, while also searching for possible signs of tidal disturbance. Our photometry reaches $\sim$3 magnitudes deeper than the discovery data, enabling robust measurements of these quantities. Both systems exhibit color–magnitude diagrams consistent with old ($\sim12.5$ Gyr), very metal-poor stellar populations. We find Gru~II to be at a distance of $\gruDKPC\pm\gruDKPCerr$ kpc, with a half-light radius of $\gruRH\arcmin\pm\gruRHerr\arcmin$ ($\gruRHPHYS\pm\gruRHPHYSerr$ pc), ellipticity $\epsilon$ = $\gruELLIP\pm\gruELLIPerr$, and absolute magnitude M$_V$ = \gruMV$\pm$\gruMVerr~mag. Hor~II is further away at a distance of $\horoDKPC^{+5.9}_{-5.5}$ kpc, and more compact with r$_h$ = $\horoRH\arcmin\pm\horoRHerr\arcmin$ ($\horoRHPHYS^{+6}_{-5}$ pc), $\epsilon$ = $\horoELLIP^{+0.20}_{-0.16}$, and M$_V$ = \horoMV$\pm$\horoMVerr~mag. Both galaxies lie within the typical size–luminosity locus of MW ultra-faint dwarfs. Gru~II shows an asymmetric morphology including multi-directional clumpy features, some of which could be suggestive of tidal disturbance. We further identify and spectroscopically confirm a new distant member just outside 3~r$_h$ in Gru~II, providing independent evidence for member stars at large projected radii. In contrast to Gru~II, Hor~II appears regular, with no significant extended structure detected to the surface-brightness limits of our data. 
\end{abstract}


\section{Introduction} \label{sec:intro}

Ultra-faint dwarf (UFD) galaxies represent the most extreme end of galaxy formation, characterized by very low luminosities ($L \lesssim 10^{5}L_{\odot}$), small stellar masses ($M_{*} \lesssim 10^{5}M_{\odot}$), and predominantly old ($\tau \sim 13.0$~Gyr), metal-poor stellar populations embedded within dark matter–dominated halos \citep[see][for a review]{Simon2019}. As such, they provide a uniquely sensitive laboratory for testing galaxy formation on the smallest scales and for probing the nature of dark matter within the Lambda Cold Dark Matter  ($\Lambda$CDM) framework \citep[e.g.,][]{Weinberg2015,Bullock2017,Strigari2018,Safarzadeh2020}.

In particular, the Milky Way (MW) hosts a rich population of UFDs that probe the lowest limits of galaxy formation. Over the past decade, wide-field optical surveys have significantly expanded the census of MW satellites, revealing a growing population of systems at the threshold of detectability \citep[e.g.][]{Balbinot2013,Belokurov2014,Laevens2014,Bechtol2015, DW2015, Kim2015a,Koposov2015a,Laevens2015,Martin2015,DW2016,Torrealba2016a,Koposov2017,Torrealba2018, Koposov2018, Mau2019, Mau2020, Cerny2021, Cerny2023a, Cerny2023b, Cerny2023c, Tan2025, Cerny2025, Tan2026}. However, many of these newly discovered systems are identified in relatively shallow imaging data, often with only a sparse population of resolved stars. This limits the precision with which their fundamental properties (e.g. distance, luminosity and size) can be constrained. Deep, wide-field follow-up imaging is therefore essential to robustly characterize these systems and to enable sensitive searches for faint, extended features that may signal tidal disturbance \citep[e.g.][]{Coleman2007,Sand2009,Sand2010,Sand2012,Munoz2010,Roderick2015,Carlin2017, Mutlu-Pakdil2018, Casey2025}.

Grus~II (Gru~II) and Horologium~II (Hor~II) are two such MW UFDs discovered in Dark Energy Survey (DES) data, with Hor~II identified in the DES Y1A1 public release \citep{Kim2015a} and Gru~II in the DES year-two quick release \citep[Y2Q1;][]{DW2015}. Subsequent spectroscopic studies have provided initial constraints on their kinematics and chemical properties, though these are not yet sufficient to fully establish their dynamical states. For Gru~II, \citet{Simon2020} identified 21 member stars and measured a systemic velocity of $v_{\mathrm{hel}} = -110.0 \pm 0.5~\mathrm{km~s^{-1}}$ and a mean metallicity of $\langle \mathrm{[Fe/H]} \rangle = -2.51 \pm 0.11$, with an upper limit on the metallicity dispersion of $\sigma_{\mathrm{[Fe/H]}} < 0.45$. They were only able to place an upper limit on the velocity dispersion, $\sigma_{v} < 2.0~\mathrm{km~s^{-1}}$ at 95.5\% confidence. Nevertheless, its low metallicity and relatively large physical size strongly favor a classification as a dwarf galaxy, a conclusion further supported by the absence of significant mass segregation among its member stars \citep{Baumgardt2022}.
Gru~II also hosts one star enhanced in $r$-process elements \citep{Hansen2020}, but this does not necessarily aid in classifying the system \citep[e.g.,][]{Zaremba2025}. Hor~II, by contrast, is more sparsely populated and remains less well constrained. It exhibits properties broadly consistent with other UFDs, including a low luminosity and metal-poor stellar population \citep{Kim2015a,Richstein2024}, but its dynamical status is uncertain due to the limited number of spectroscopic members \citep[2 likely and 1 candidate;][]{Fritz2019}. Independent analyses based on stellar mass segregation nevertheless suggest that it is also likely a dwarf galaxy \citep{Baumgardt2022}. Together, these systems lie in a regime where their nature remains ambiguous, underscoring the need for deeper photometric and spectroscopic follow-up.

A key open question for both systems is whether they exhibit signs of tidal disruption or are susceptible to it. \citet{Simon2020} show that even under conservative assumptions, the inferred tidal radius of Gru~II is only a few times its half-light radius (hereafter, $r_{h}$), suggesting that a non-negligible fraction of its stars may be vulnerable to stripping. Similarly, \citet{Pace2022} argue, based on orbital considerations and its low mean density relative to the MW at pericenter, that Gru~II may be undergoing tidal disruption, consistent with broader expectations for satellites on such orbits \citep[e.g.,][]{Battaglia2022}. Identifying and characterizing tidal features is critical, as the presence of unbound stars may bias  dynamical measurements that are derived under the assumption of dynamical equilibrium. 

In this context, deep imaging combined with complementary techniques offers a powerful approach. In particular, narrow-band photometric surveys such as Mapping the Ancient Galaxy In CaHK (MAGIC) survey (NOIRLab Prop. ID 2023B-646244; \citealt{Chiti2026b}) in the Southern Hemisphere and the Pristine survey \citep{Starkenburg2017} in the Northern Hemisphere, enable the estimation of photometric metallicities using Ca~II H\&K-sensitive filters, providing an efficient means to identify extremely metal-poor candidate members in the outskirts of dwarf galaxies. When combined with color–magnitude and astrometric selections, this approach effectively recovers stellar populations across extended tidal features, as demonstrated in the case of Bo\"otes~I \citep{Longeard2022}, Hercules \citep{Longeard2023}, Crater~II \citep{Atzberger2026} and Pictor~II \citep{Chiti2026a}. By incorporating astrometric constraints from the \textit{Gaia} mission \citep{Gaia2016,Gaia2023}, such selections can isolate high-probability members at large radii, enabling targeted spectroscopic follow-up to test for the presence of extended or tidally stripped populations. 

In this paper, we present deep, wide-field Magellan/Megacam imaging of Gru~II and Hor~II, reaching several magnitudes deeper than the discovery data. Using these observations, we derive improved constraints on the distances, structural parameters, and luminosities of both systems, and search for faint stellar extensions through a matched-filter technique. In the case of Gru~II, we further incorporate photometric metallicity information to identify candidate members in the outskirts, which we follow up spectroscopically to assess their association with the system. In Section~\ref{sec:data}, we describe the observations and data reduction. In Section~\ref{sec:analysis}, we present the analysis, including color-magnitude diagrams (CMDs), distance determination, structural parameter and luminosity estimates and the search for extended features. In Section~\ref{sec:discussion}, we discuss the implications of our results in the context of tidal disruption and the nature of UFDs. Finally, we summarize the main results in Section~\ref{sec:summary}.

\section{Observations and Data Reduction} \label{sec:data}

\subsection{Imaging Observations} \label{subsec:imaging}
\begin{figure*}[t]
    \centering
    \includegraphics[width=0.425\textwidth]{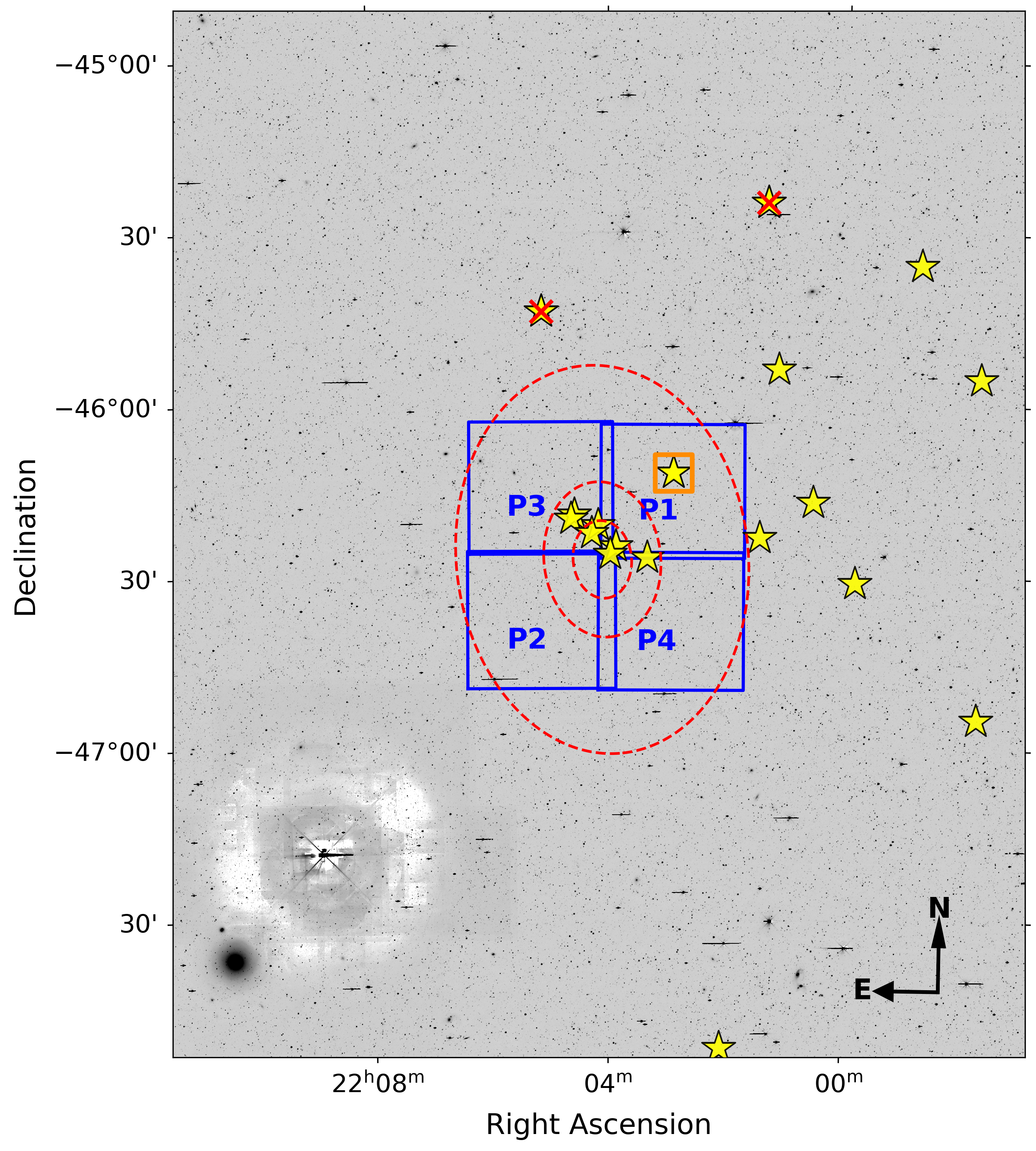}
    \includegraphics[width=0.51\textwidth]{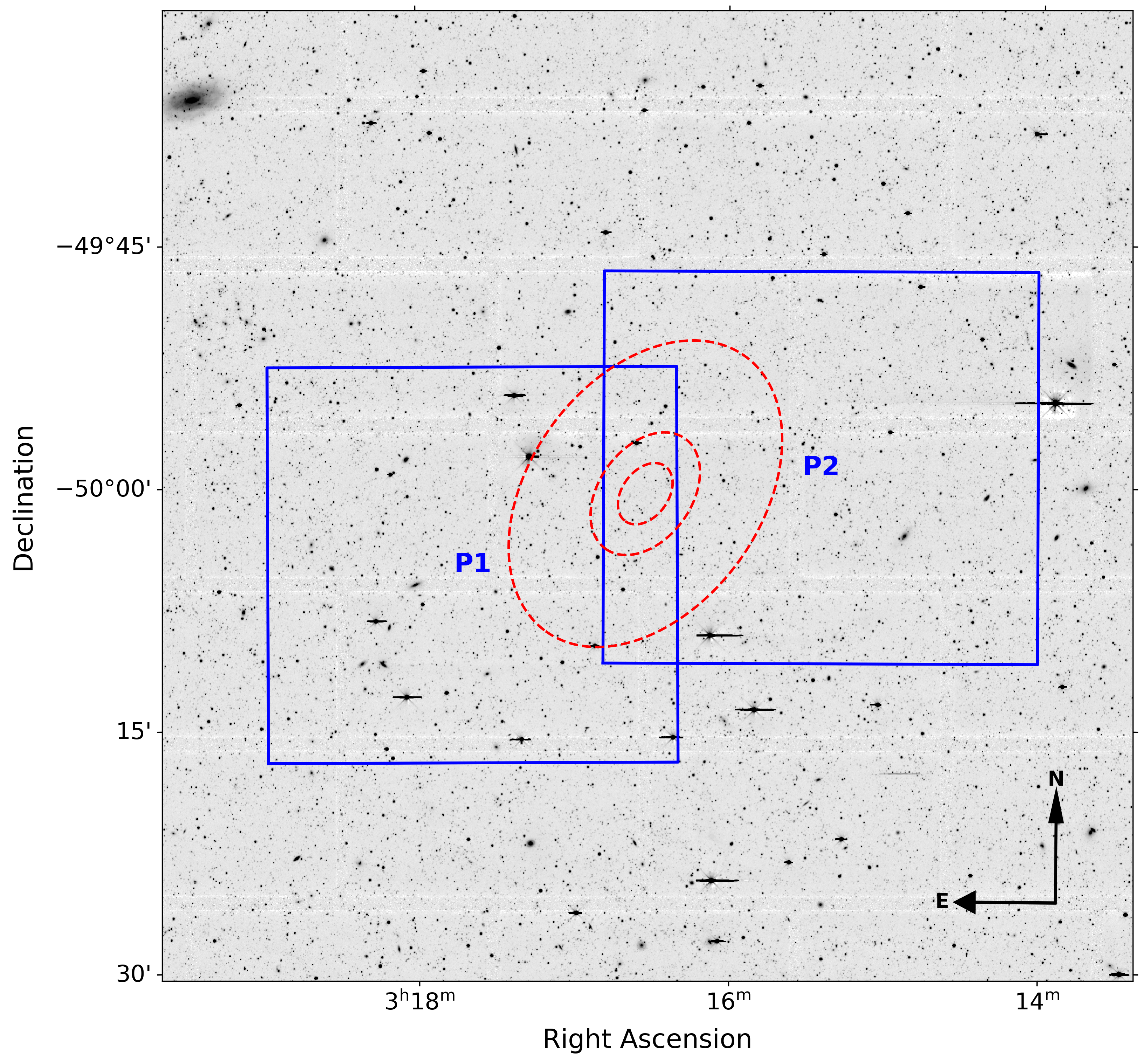} 
    
    \caption{Outline of our Megacam pointings for Gru~II (left) and Hor~II (right) on the backdrop of DESI Legacy Imaging Survey DR10 $i$-band images. The prominent feature in the lower left of the Gru~II panel is a bright star artifact in the background image and is not related to the satellite. In the Gru~II field, the 18 member candidates with [Fe/H]$_{\text{MAGIC}}$~$< -2.0$ and $g < 20.75$ discussed in Section~\ref{subsec:spectra} are shown as yellow stars. Additionally, from the three observed Magellan/MagE spectroscopic targets, the new confirmed member is highlighted using an orange box and two non-members are marked by red crosses. The ellipses corresponding to 1, 2 and 5 times the half-light radius for each satellite (derived in Section~\ref{subsec:struct}), are shown using red dashed outlines.}
    \label{fig:pointings}
\end{figure*}

Gru~II and Hor~II were observed with multiple pointings %(PI: Mutlu-Pakdil, B.) 
using the Megacam instrument \citep{McLeod2015} mounted at the $f$/5 focus of the Magellan Clay telescope. Magellan/Megacam is made up of a mosaic of 36 CCDs, each having 2048$\times$4608 pixels; however, two CCDs in the mosaic were non-functional at the time of the observations. We used $2\times2$ binning for our observations, resulting in a plate scale of 0.16\arcsec/pixel. Gru~II was imaged in four pointings and Hor~II in two, covering approximate fields of view (FOVs) of $\sim$~48\arcmin$\times$48\arcmin\ and 48\arcmin$\times$30\arcmin, respectively, as shown in Figure~\ref{fig:pointings}. For every pointing, we obtained multiple exposures of $300$~s each in the $g$- and $r$-bands (see Table~\ref{table:1}).

\begin{table*}[t]
\caption{Summary of Megacam observations and field completeness.}
\centering
\begin{tabular*}{\linewidth}{c c c c c c c c c c}
 \hline \hline
 Dwarf Name & Pointing & R.A. & Dec. & UT Date & Filter & Exp. time & Seeing & \multicolumn{2}{c}{Completeness} \\ 
            &          & (hh:mm:ss) & (dd:mm:ss) & & & (s) & (arcsec) & 50\% (mag) & 90\% (mag) \\
 \hline
  \multirow{8}{*}{Gru~II} & P1 & 22:02:54.367 & $-$46:15:01.099 & 2022/10/21 & $g$ & 7$\times$300 & 0.83 & 27.25 & 26.45 \\
                           &    & & & 2022/10/22 & $r$ & 6$\times$300 & 0.60 & 26.74 & 25.91 \\
                           & P2 & 22:05:08.338 & $-$46:37:36.400 & 2022/10/22 & $g$ & 7$\times$300 & 0.66 & 27.06 & 26.35 \\
                           &    & & & 2022/10/23 & $r$ & 6$\times$300 & 0.69 & 26.61 & 25.63 \\
                           & P3 & 22:05:08.400 & $-$46:14:46.198 & 2022/10/23 & $g$ & 6$\times$300 & 0.63 & 27.11 & 26.41 \\
                           &    & & & 2022/10/23 & $r$ & 7$\times$300 & 0.69 & 26.69 & 25.60 \\
                           & P4 & 22:02:52.897 & $-$46:37:46.205 & 2022/10/25 & $g$ & 6$\times$300 & 0.59 & 27.16 & 26.43 \\
                           &    & & & 2022/10/25 & $r$ & 6$\times$300 & 0.53 & 26.68 & 25.79 \\
 \hline \\
  \multirow{4}{*}{Hor~II} & P1 & 3:17:39.410 & $-$50:04:59.693 & 2022/10/27 & $g$ & 12$\times$300 & 0.92 & 27.08 & 26.16 \\
                           &    & & & 2022/10/27 & $r$ & 12$\times$300 & 0.80 & 26.59 & 25.55 \\
                           & P2 & 3:15:24.520 & $-$49:58:48.603 & 2022/10/28 & $g$ & 6$\times$300 & 0.85 & 26.86 & 26.10 \\
                           &    & & & 2022/10/28 & $r$ & 6$\times$300 & 0.62 & 26.42 & 25.44 \\
  \hline \\
\end{tabular*}
\label{table:1}
\end{table*}

Data reduction was carried out using the Megacam pipeline developed at the Harvard-Smithsonian Center for Astrophysics by M. Conroy, J. Roll, and B. McLeod. The pipeline performs standard image reduction procedures, including bias subtraction, flat-field correction, cosmic-ray removal, and astrometric calibration. The individual dithered frames were then coadded using {\fontfamily{cmtt}\selectfont SWarp} \citep{Bertin2002}. The affected regions corresponding to the non-functional chips were masked and excluded in our subsequent analysis.

We performed point-spread function (PSF) photometry on each coadded image using the DAOPHOT~II/ALLSTAR package \citep{Stetson1987,Stetson1994}, and followed the methodology detailed in \citet{Sand2009}, and \citet{Mutlu-Pakdil2018}. Briefly, we adopted a quadratically varying PSF across the FOV to create the model PSF. We ran ALLSTAR two times: first, on the final stacked image, and then on the image obtained after subtracting the stars found in the first run. This methodology was adopted so as to recover much fainter sources. The source catalogs from the $g$- and $r$-band images were then matched using DAOMATCH/DAOMASTER \citep{Stetson1993}, and a final round of photometry was performed simultaneously in both bands with ALLFRAME \citep{Stetson1994}, allowing the detection and measurement of fainter sources than would be possible from the individual images alone. We then culled our catalog by removing objects that were not consistent with well-measured point sources. Specifically, we applied cuts to the $\chi^{2}$, photometric uncertainty, and sharpness distributions as a function of source magnitude. The selection boundaries were defined using analytic functions with coefficients determined empirically for each pointing and filter to trace the loci of well-measured sources. The allowed ranges were broadened toward fainter magnitudes to account for the increasing photometric scatter. We positionally matched our source catalogs derived from the $g$- and $r$-band images using a maximum matching radius of 0.5\arcsec. Only point sources detected in both bands were used to create the catalog corresponding to each pointing. 

We calibrated our photometry corresponding to individual pointings of both galaxies by matching to the DESI Legacy Imaging Surveys Data Release 10 \citep[DR10, ][]{Dey2019} catalog. The DR10 photometry is on the native AB system of the Dark Energy Camera (DECam) instrument at the 4~m V\`ictor M. Blanco Telescope on Cerro Tololo, Chile, with zero-points tied to PS1 DR1 \citep{Dey2019}. For calibration, we used all the stars within each pointing's FOV where 18 $< g <$ 20.5 and 18 $< r <$ 20.5. We then applied a correction to the magnitudes for Galactic extinction using the \citet{Schlegel1998} dust maps, adopting the extinction coefficients of $A_g / E(B-V) = 3.237$ and $A_r / E(B-V) = 2.176$  from \citet{Schlafly2011}  appropriate for the DECam $g$, $r$ bandpasses. In the following sections of the paper, we use extinction-corrected photometry, unless otherwise stated.

In order to estimate the photometric uncertainties and completeness as a function of magnitude and color for each pointing, we carried out artificial star tests using the ADDSTAR module of DAOPHOT. We followed the same methodology adopted by various works in the literature \citep[][]{Sand2009, Sand2012, Mutlu-Pakdil2018, Casey2025}. Briefly, artificial stars were inserted into our images on a regular grid. The injected stars' $r$-band magnitudes were chosen randomly within the range 18 to 29 mag such that fainter magnitudes have exponentially increasing probabilities. Subsequently, each star was assigned a random $g - r$ color within the range $-0.5$ to $1.5$ mag. The injection procedure was repeated ten times, each run adding $\sim$100,000 artificial stars to each field. Next, we carried out photometry on the artificial images using DAOPHOT/ALLSTAR in the same manner as for the observed data and applied the same criteria for selecting point-sources (chi square, magnitude, magnitude error, and sharpness) in the artificial star catalogs. Finally, the 50\% and 90\% completeness (reported in Table~\ref{table:1}) and photometric uncertainties for each pointing were estimated from these artificial star catalogs.

As the final step in the catalog creation for each dwarf, we merged the catalogs from the individual pointings while ensuring that sources in the overlapping regions were included only once in the final catalog. In overlap regions, we retained detections from the deepest available pointing based on the $r$-band 50\% completeness limits (see Table~\ref{table:1}), using the corresponding weight maps to account for faulty CCD chips. We present the final photometric catalogs for every point source in our Gru~II and Hor~II fields in Tables \ref{tab:gru_table} and \ref{tab:horo_table}, respectively. Table columns include the equatorial coordinates, calibrated magnitudes (not corrected for extinction), DAOPHOT photometric uncertainties, and the Galactic extinction.

\begin{table*}[t]
\caption{\textbf{Gru~II Photometry in the DECaLS/DECam Photometric System}}
    %\centering
    \begin{tabular}{ccccccccc}
    \hline \hline
        Star No. & R.A. & Dec. & $g$ & $\sigma_g$ & $A_{g}$ & $r$ & $\sigma_r$ & $A_{r}$ \\
         & (deg J2000.0) & (deg J2000.0) & (mag) & (mag) & (mag) & (mag) & (mag) & (mag) \\
        \hline
        0 & 330.422850 & $-$46.427809 & 26.56 & 0.25 & 0.06 & 26.30 & 0.28 & 0.04 \\
        1 & 330.422860 & $-$46.429135 & 26.29 & 0.15 & 0.06 & 26.40 & 0.31 & 0.04 \\
        2 & 330.422870 & $-$46.420579 & 25.69 & 0.24 & 0.06 & 25.26 & 0.16 & 0.04 \\
        3 & 330.422940 & $-$46.444826 & 25.13 & 0.03 & 0.06 & 24.94 & 0.07 & 0.04 \\
        \hline
    \end{tabular}
    \label{tab:gru_table}
    \\
    (This table is available in its entirety in a machine-readable form in the online journal. A portion is shown here for guidance regarding its form and content.)
\end{table*}

\begin{table*}[t]
\caption{\textbf{Hor~II Photometry in the DECaLS/DECam Photometric System}}
    %\centering
    \begin{tabular}{ccccccccc}
    \hline \hline
        Star No. & R.A. & Dec. & $g$ & $\sigma_g$ & $A_{g}$ & $r$ & $\sigma_r$ & $A_{r}$ \\
         & (deg J2000.0) & (deg J2000.0) & (mag) & (mag) & (mag) & (mag) & (mag) & (mag) \\
        \hline
        0 & 48.503587 & $-$50.166618 & 25.46 & 0.15 & 0.07 & 24.42 & 0.07 & 0.05 \\
        1 & 48.503614 & $-$50.166773 & 25.24 & 0.11 & 0.07 & 24.28 & 0.06 & 0.05 \\
        2 & 48.503742 & $-$50.145949 & 26.60 & 0.26 & 0.07 & 26.74 & 0.35 & 0.04 \\
        3 & 48.503743 & $-$50.134197 & 26.02 & 0.36 & 0.07 & 23.86 & 0.06 & 0.04 \\
        \hline
    \end{tabular}
    \label{tab:horo_table}
    \\
    (This table is available in its entirety in a machine-readable form in the online journal. A portion is shown here for guidance regarding its form and content.)    
\end{table*}

\subsection{Spectroscopic Selection \& Observations} \label{subsec:spectra}

As a complementary avenue to assess whether the Gru~II UFD hosts extended features, we obtained medium-resolution ($R\sim6,000$) spectra of three stars that are candidate members of the system in its outskirts ($>$~3~$r_{h}$) using the Magellan/MagE spectrograph \citep{Marshall2008}. The stars were observed with the 0.7\arcsec\ slit in good seeing conditions ($\sim$0.5\arcsec) on October 19, 2025. The selection of these candidates follows previous searches for low metallicity stars in the outskirts of UFDs \citep[e.g., ][]{Chiti2021b, Longeard2022}. Specifically, these stars were selected to a) be along the red giant branch (RGB) of Gru~II, b) have proper motions consistent within 3\,$\sigma$ with that of the system ($\mu_{\alpha} \cos \delta$ = 0.384$\pm$0.033~mas~yr$^{-1}$ and $\mu_{\delta}$ = $-$1.484$\pm$0.040~mas~yr$^{-1}$, \citealt{Pace2022}) based on \textit{Gaia} DR3 \citep{Gaia2016,Gaia2023}, c) be consistent with a non-detection of a parallax ($\varpi - 3\sigma_{\varpi} < 0$), d) not be in the \textit{Gaia} DR3 variable star catalog\footnote{\url{https://gea.esac.esa.int/archive/}, \texttt{gaiadr3.vari\_summary}} \citep{Eyler2023} and e) have low photometric metallicity estimates based on MAGIC survey photometry. 

\begin{figure*}[t]
    \centering
    \includegraphics[width=1.0\textwidth]{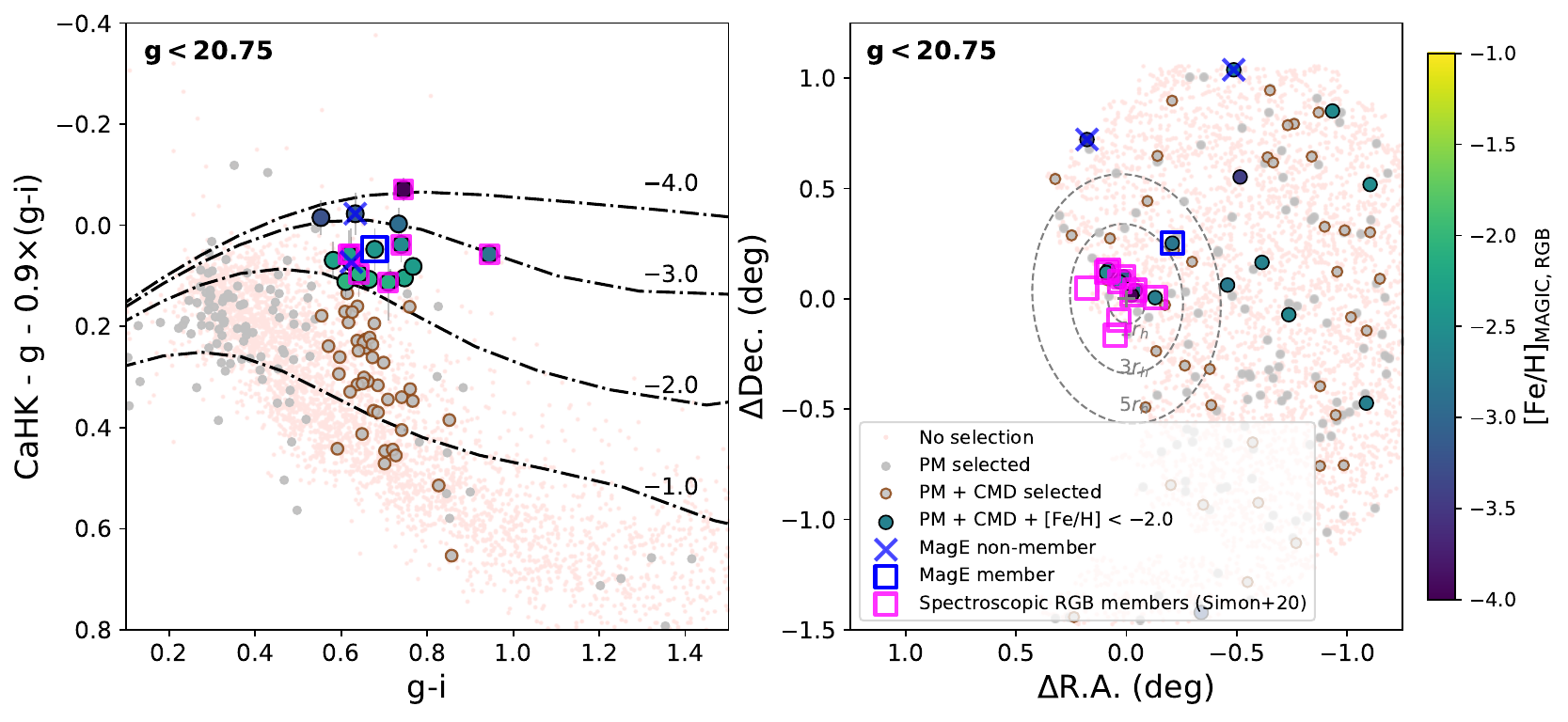}
    \caption{Left: Metallicity-sensitive CaHK color-color plot of stars within 1.5$^\circ$\ of the center of Gru~II (see Section~\ref{subsec:spectra}).
   Only stars with $g < 20.75$ are shown, corresponding to our target selection for MagE spectroscopy. Sources in the catalog without applying any selection criteria are shown as rose-colored points. Gray circles indicate sources with \textit{Gaia} DR3 proper motions consistent with membership, while those outlined in brown additionally pass the CMD-based selection. Sources that further satisfy [Fe/H]$_{\mathrm{MAGIC}} < -2.0$ are shown as large circles and are colored by their MAGIC metallicities. Spectroscopic RGB members of Gru~II in \citet{Simon2020} that also have $g <20.75$ are shown (magenta squares), demonstrating that Gru~II members generally separate from foreground stars based on their MAGIC photometric metallicities. The stars followed-up in this work with MagE (see Table~\ref{tab:spectable}) are indicated with blue crosses or a square. Overplotted contours indicate model predictions for where stars at various metallicities would lie at $\log\,g = 3.0$. Right: Spatial distribution of the same samples as shown in the left panel. The 1, 3, and 5 times the $r_{h}$ of Gru~II are shown as gray dashed lines. We note that the presence of the OC stream likely contaminates the region (see Section~\ref{sec:discussion}).
    }
    \label{fig:cahkselection_GruII}
\end{figure*}

To expand on the latter, the MAGIC survey is obtaining metallicity-sensitive narrow-band photometry using the DECam instrument over the Ca II H\&K lines across a number of dwarf galaxies in the southern hemisphere \citep{Chiti2026b}. 
This narrow-band data is coupled with broadband $g, r, i$ information from the DECam Local Volume Exploration survey Data Release 2 \citep[DELVE DR2; ][]{delvedr2} to derive photometric metallicities.
The photometric metallicities for MAGIC are generated using a generalization of the techniques presented in \citet{Chiti2020, Chiti2021b}, and further detailed in \citet{Chiti2026b}.
Candidates were required to have [Fe/H]$_{\rm MAGIC} < -2.0$, motivated by the previously measured very metal-poor stellar population of Gru~II \citep[e.g.,][]{Simon2020}. This threshold was primarily chosen to optimize sample purity by reducing contamination from the generally more metal-rich foreground MW disk and halo populations \citep{Conroy2019,Youakim2020}. We restricted our sample to stars with $g < 20.75$ to enable efficient spectroscopic follow-up, and to those located within $\sim$1.5$^\circ$ of the center of Gru~II, corresponding to roughly 15 times $r_{h}$ (see Section~\ref{subsec:struct}), in order to encompass potential extended structure while limiting contamination from more distant foreground and background sources. There were 18 such sources and these are indicated using yellow star symbols in the left panel of Fig.~\ref{fig:pointings} and in the right panel of Fig.~\ref{fig:cahkselection_GruII}, including multiple spectroscopically confirmed members in the center of Gru~II \citep{Simon2020}. We note here that 8 of these candidates fall within our deep imaging FOV.

The identification of these sources is further illustrated in Fig.~\ref{fig:cahkselection_GruII}, where we show our selection of candidate members in the vicinity of Gru~II using MAGIC photometry. Specifically, the color-color space defined by CaHK$- g - 0.9\times(g-i)$ vs. $(g-i)$ has been shown to separate metal-rich from metal-poor stars \citep[e.g.,][]{Keller2007, Starkenburg2017}. Since UFD members are generally more metal-poor than ambient MW halo stars, they separate from foreground stars in this color-color space, enabling their efficient identification and spectroscopic follow-up. This is illustrated in the left panel of Fig.~\ref{fig:cahkselection_GruII}, which shows stars in the MAGIC catalog with $g < 20.75$ as rose-colored points, those having proper motions consistent ($<$~3$\sigma$) with membership in Gru~II as gray circles, and those that additionally pass the CMD selection as gray circles with a brown outline. Sources that further satisfy [Fe/H]$_{\mathrm{MAGIC}} < -2.0$ are highlighted as large circles and are colored by their MAGIC metallicities. Spectroscopically confirmed RGB members of Gru~II from \citet{Simon2020} with $g < 20.75$ are highlighted using magenta squares. These stars occupy a distinct very metal-poor region ([Fe/H] $\lesssim -2.0$) in the color-color space relative to the bulk of field stars, and our selection recovers all such members apart from those outside the catalog coverage. For completeness, we note that the MAGIC coverage out to 1.5$^\circ$ of Gru~II contains $\sim9700$ stars with $g < 20.75$. Applying the CMD selection reduces this to 956 sources, while the subsequent proper-motion and astrometric selections yield 60 candidates. Finally, 18 sources satisfy the photometric metallicity criterion [Fe/H]$_{\mathrm{MAGIC}} < -2.0$.

We do note two caveats that limit the utility of MAGIC photometry in the vicinity of Gru~II. 
First, the catalog coverage is absent eastward of the center of Gru~II (see right panel of Fig.~\ref{fig:cahkselection_GruII}); this absence is due to the bright, heavily saturated star artifact seen in the bottom left of Fig.~\ref{fig:pointings}.
This star was captured by the large DECam field-of-view ($\sim$2.2$^\circ$ diameter) within the MAGIC survey pointing eastward of Gru~II, and unfortunately rendered the entire exposure unusable. 
This issue does not apply to the deep Megacam pointings (P1-P4) from which structural parameters are derived, as none of these pointings encompassed the bright star.
Additionally, the Orphan/Chenab (OC) stream appears westward of Gru~II \citep[see Sections~\ref{subsec:cmds} and \ref{sec:discussion}]{Grillmair2006, Belokurov2007, Shipp2018} and is likely the source of the low metallicity candidates westward of the system.
Accordingly, we limited MagE spectroscopic follow-up to just three distant candidates: one candidate within 5\,$r_h$ of the system, and two others $>5\,r_h$ northward of the system.
Information on these three distant Gru~II candidates is shown in Table~\ref{tab:spectable}. 

The data were reduced using the standard reduction techniques in the CarPy pipeline~\citep{Kelson2003}.
Radial velocities were derived following \citet{Chiti2026a}, in which the H$\alpha$ region of the spectra were cross-correlated with a MagE spectrum of HD122563 placed at $-25.61$\,km\,s$^{-1}$, observed with the same instrument setup.
From this, we found that only one star (\textit{Gaia} DR3 Source ID: 6561615567541856640) had a radial velocity ($-$106.3 $\pm$ 5.1~km s$^{-1}$) consistent with membership of the system (v$_{\text{sys}} = -110.0 \pm 0.5$; \citealt{Simon2020}) within a 3\,$\sigma$ selection. This source is highlighted using an orange square box in the left panel of Fig.~\ref{fig:pointings}. 
We derived the metallicity of this star using the well-established KP calibration \citep{Beers1999}, where the pseudo-equivalent width of the Ca II K line at 3933\,{\AA} along with the $B-V$ color of the star is mapped to a metallicity. 
We used a 12~Gyr, [Fe/H] = $-2.5$ Dartmouth isochrone \citep{Dotter2008} to convert from a DECam $g-i$ color to $B-V$ color, and then followed the implementation detailed in \citet{Chiti2018} to derive the metallicity. 
The resulting metallicity is \feh = $-2.69 \pm 0.30$, consistent with being a member of Gru~II. The two non-members from our analysis are marked with red crosses in Fig.~\ref{fig:pointings}, left panel\footnote{We note here that our original selection for Gru~II candidates did not have a \textit{Gaia} DR3 \texttt{ruwe} or \texttt{astrometric\_excess\_noise} cut (see Section~\ref{subsec:gaia_members}).
However, all the selected Gru~II candidates do have \texttt{ruwe}~$<~1.4$ and only one does not have \texttt{astrometric\_noise\_excess}~$~<~2$; the latter is a spectroscopic non-member (\textit{Gaia} DR3 source ID: 6561720364744285696)}.

\begin{table*}[tb!]
\caption{Magellan/MagE Spectroscopic Observations of Gru~II}\label{tab:spectable}
\begin{minipage}[b]{0.95\linewidth}\centering
\begin{tabular}{lcccccccc}
\tablewidth{0pt}
\hline
\hline
\textit{Gaia} DR3 Source ID & R.A. (h:m:s) & Dec (d:m:s) & $g$ & $t_{\text{exp}}$ & [Fe/H]$_{\text{MAGIC}}$ & $v_{\text{helio}}$ & Member\\
 {} & (J2000) & (J2000) &  (mag) & (min) & (dex) & (km s$^{-1}$) & \\
\hline
6561720364744285696 & 22:01:19.66 & $-$45:24:38.70 & 20.74 & 45 & $-2.22\pm0.33$ & $-$22.5 $\pm$ 6.0 & No\\
6567466378151096576 & 22:05:07.20	& $-$45:43:43.02 & 20.52 & 45 & $-3.15\pm0.51$ & $-$47.1 $\pm$ 4.7 & No\\
6561615567541856640 & 22:02:53.79	& $-$46:11:53.76 &  20.61 & 90 & $-2.43\pm0.32$ & $-$106.3 $\pm$ 5.1 & Yes\\

\hline

\end{tabular}
\begin{tablenotes}
\small
\item Cols. (1)--(8) list the \textit{Gaia}~DR3 Source ID, Right Ascension, Declination, $g$-band magnitude from DELVE/MAGIC, exposure time for the MagE observations, MAGIC photometric metallicity, heliocentric radial velocity measured from the MagE spectra, and Gru~II membership classification, respectively.

\end{tablenotes}
\end{minipage}    
\end{table*}

\begin{figure}[t]
    \centering
    \includegraphics[width=0.48\textwidth]{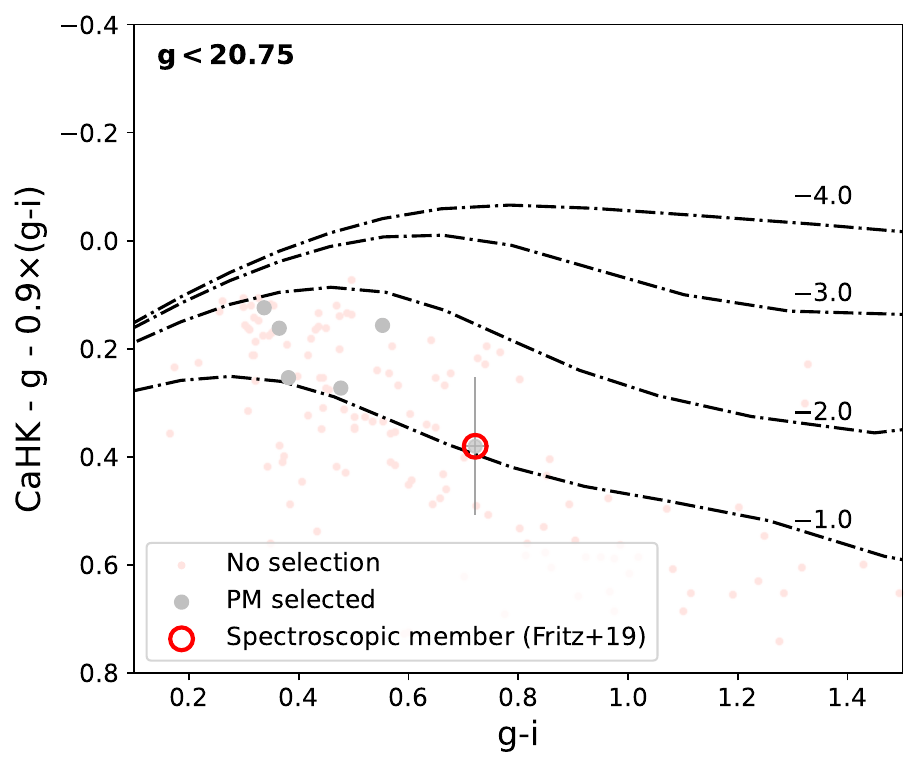}
    \caption{Similar to the left panel of Figure~\ref{fig:cahkselection_GruII}, but adapted for Hor~II.
    Specifically, we show a CaHK color-color plot of stars within 10\,$r_h$ of the center of Hor~II, using CaHK photometry (see Section~\ref{subsec:spectra}). Sources in the catalog without applying any selection criteria are shown as rose-colored points. Gray circles indicate sources with \textit{Gaia} DR3 proper motions consistent with membership. However, none of these sources pass our CMD-based selection. We show the only spectroscopic RGB member in \citet{Fritz2019} that we recover, which has relatively high photometric metallicity, albeit with large uncertainty ([Fe/H]$_{\text{MAGIC}} = -1.15\pm0.60$). The potential \textit{Gaia} DR3 Hor~II member discussed in Section~\ref{subsec:gaia_members} does not have an entry in the MAGIC catalog. Overplotted contours are same as those in the left panel of Fig.~\ref{fig:cahkselection_GruII}.}
    \label{fig:cahkselection_HorII}
\end{figure}

We note for completeness that while the most recent version\footnote{As of 04/18/2026} of the MAGIC catalog includes undithered coverage of the Hor~II field, it does not yield any viable candidate members with our selection criteria (see Figure~\ref{fig:cahkselection_HorII}).
We therefore did not pursue spectroscopic follow-up for Hor~II. However, we do note that one confirmed member in the spectroscopic catalog of \citet{Fritz2019}, designated \texttt{horo2\_8\_156}, is included in the MAGIC catalog. This source has a MAGIC metallicity generally too high to be consistent with UFD membership, albeit with large uncertainty ([Fe/H]$_{\text{MAGIC}} = -1.15\pm0.60$).
Nevertheless, for future follow-up studies, we present a search for candidate members in this UFD based solely on our deep Megacam photometry and \textit{Gaia}~DR3 astrometry in Section~\ref{subsec:gaia_members}.
The candidate Hor~II star discussed in that latter section has no entry in the MAGIC catalog due to incompleteness from the undithered coverage, precluding a statement on whether its MAGIC metallicity is consistent with membership.

\section{Analysis}\label{sec:analysis}
\subsection{Color-Magnitude Diagrams} \label{subsec:cmds}
\begin{figure*}[t]
    \centering
    \includegraphics[width=0.49\textwidth]{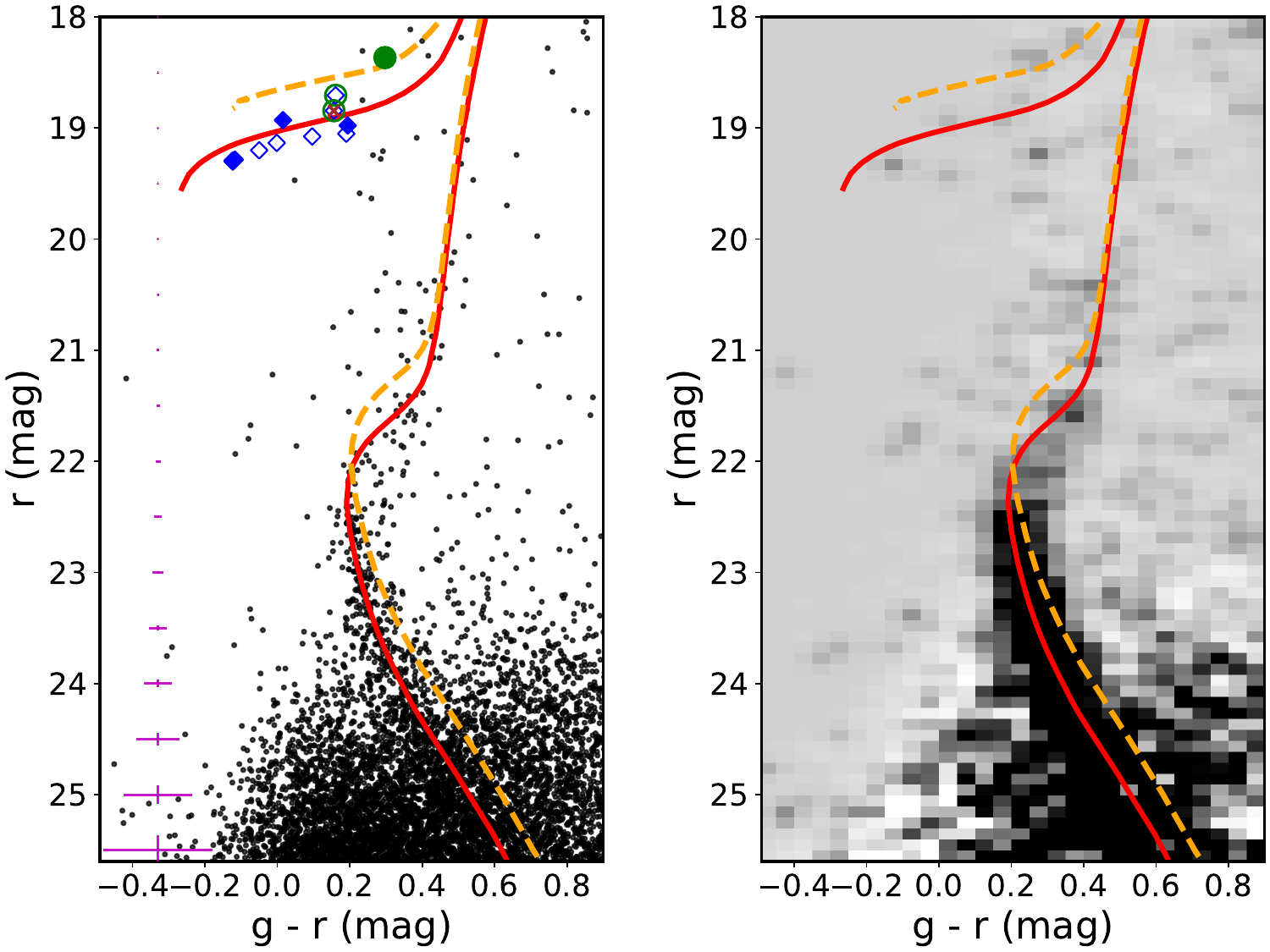}
    \includegraphics[width=0.49\textwidth]{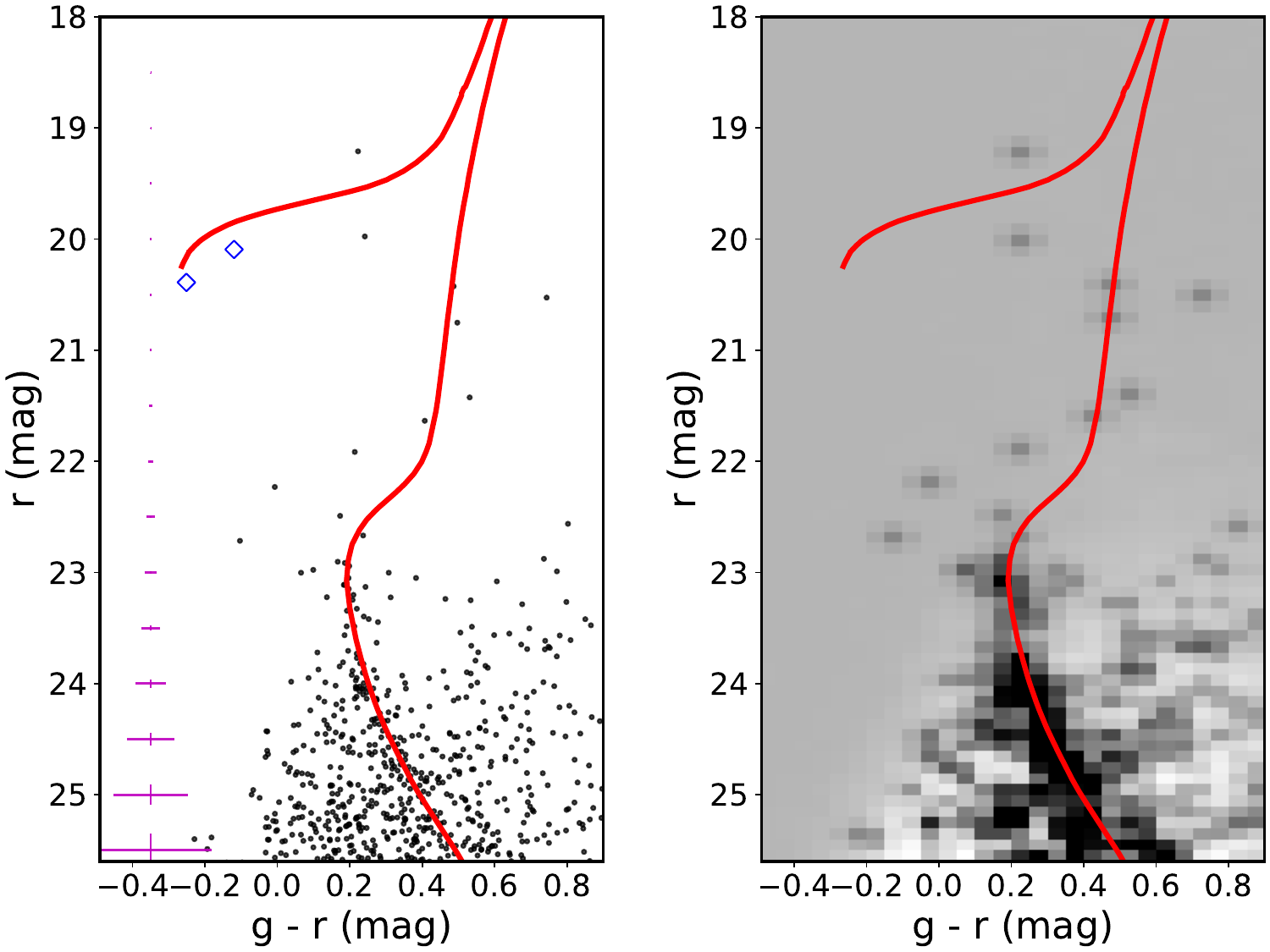} 
    
    \hfill\parbox[t]{0.24\textwidth}{(a) Gru~II}\hfill
    \parbox[t]{0.24\textwidth}{(b) Hor~II}\hfill
    
    \caption{CMD and Hess diagrams for Gru~II and Hor~II including sources within their respective $r_{h}$ (Section~\ref{subsec:struct}). On the CMDs, candidate HB stars within 3~$r_{h}$ are shown as blue open diamonds, with those within 1~$r_{h}$ as blue filled diamonds. %(note that Horo~II has only one HB candidate within three times its half-light radius).
    Among the seven RR~Lyrae stars identified around Gru~II by \citet{Martinez-Vazquez2019} and \citet{Tau2024}, the three lying within our FOV are shown as green filled (within 1~$r_{h}$) and open (outside 1~$r_{h}$) circles. The likely Gru~II member is marked with a red cross, while the two brighter ones may be associated with the OC stream \citep{Martinez-Vazquez2019}. The magenta error bars indicate the mean color and magnitude uncertainties. The red curve shows a PARSEC isochrone of age 12.5~Gyr and metallicity $\mathrm{[Fe/H]}=-2.2$. The orange dashed curve (panel a)  corresponds to a PARSEC isochrone of age 12.5~Gyr and $\mathrm{[Fe/H]}=-1.9$, shifted to the OC stream distance to guide the identification of potential stream contaminants. The Hess diagrams on the right for each system are background-subtracted and binned as described in the main text.}
    \label{fig:cmds}
\end{figure*}

In the left panels of Figure \ref{fig:cmds} (a) and (b), we show the CMDs of Gru~II and Hor~II including sources within their respective $r_{h}$ (see Section~\ref{subsec:struct}). The photometric uncertainties in magnitude and color as determined from the artificial star tests are shown as magenta error bars at an arbitrary color. In red is the PARSEC isochrone \citep{Bressan2012} with age 12.5~Gyr, and metallicity, $\text{[Fe/H]}=-2.2$, which corresponds to the best-fit isochrone from Section~\ref{subsec:distance}. Our CMDs extend to $\sim$3.5~mag below the main-sequence turnoff (MSTO) for Gru~II and $\sim$2.5~mag below the MSTO for Hor~II. While the \textit{HST} observations of \citet{Richstein2024} reach approximately 0.5~mag deeper below the turnoff in Gru~II and $\sim$1~mag deeper in Hor~II, the depth achieved by our Megacam imaging robustly traces the faint stellar populations of both systems, while providing substantially wider spatial coverage.

Gru~II is known to overlap spatially and in proper motion space with the OC stellar stream \citep{Koposov2019, Koposov2023, Shipp2019}. However, despite this projected and kinematic overlap, Gru~II differs significantly from the OC stream in radial velocity \citep[by $\sim$~90~km~s$^{-1}$;][]{Simon2020} and distance, with the OC stream located at $m-M=18.17$~mag \citep[$\sim$~10~kpc closer than Gru~II;][]{Martinez-Vazquez2019}. The OC stream has a mean metallicity of $\mathrm{[Fe/H]} = -1.9$ \citep{Koposov2023, Hawkins2023}. To guide the identification of potential stream contaminants, we therefore overplot in Figure~\ref{fig:cmds} (a), an additional PARSEC isochrone (orange dashed curve) with age 12.5~Gyr and $\mathrm{[Fe/H]} = -1.9$, shifted to the stream distance. 

Candidate horizontal branch (HB) stars within 3~$r_{h}$ of each galaxy are shown as blue open diamonds (11 in Gru~II and 2 in Hor~II) and those within 1~$r_{h}$ are highlighted as blue filled diamonds (4 in Gru~II and none in Hor~II). These candidates are selected such that $g-r < 0.2$ and their $r$ magnitudes lie within $\pm0.2$ mag of the HB sequence defined by the PARSEC isochrone. A total of seven RR~Lyrae stars have been detected in the vicinity of Gru~II, although only two are considered members of the UFD by \citet{Martinez-Vazquez2019} and \citet{Tau2024}. The RR~Lyrae stars within our Gru~II field are shown as green filled circles (within 1~$r_{h}$) and green open circles (outside 1~$r_{h}$). However, only one among these is considered a member of Gru~II by \citet{Martinez-Vazquez2019} and this source is marked with a red cross in the left panel of Fig.~\ref{fig:cmds}~(a). The other two brighter RR~Lyrae stars within our Gru~II FOV are likely members of the OC stream, as suggested by \citet{Martinez-Vazquez2019}. 

The right panels for each UFD in Figure~\ref{fig:cmds} show the background-subtracted binned Hess diagrams, accentuating the number density of sources across the CMD. We derived the background from regions located beyond 3~$r_{h}$ of each UFD, well outside their main bodies.

\subsection{Distance} \label{subsec:distance}

We estimate the distance moduli of the satellites by comparing their CMDs with different theoretical isochrones, similar to the methodology adopted by \citet[][see also \citealt{Walsh2008, Mutlu-Pakdil2018, Casey2025}]{Sand2009}. For Gru~II, we use two metal-poor PARSEC isochrones of age 12.5~Gyr with \feh\ $=-2.2$ (the most metal-poor available) and $-2.0$. For Hor~II, we adopt the same two isochrones along with an additional 13.5~Gyr, \feh\ $=-2.1$ isochrone, as used in the discovery paper by \citet{Kim2015a}.

To derive the distance modulus for Gru~II, we choose all sources with $r$ $<=$ 24~mag within 1~$r_{h}$. We vary the distance modulus in steps of 0.025 mag from 17.5 to 19.5 mag, in line with the estimate by \citet[][]{Martinez-Vazquez2019}, $m-M$ = 18.71$\pm$0.10~mag. For each value, we count the number of sources consistent with the isochrone, within a tubular region defined by the photometric uncertainties. We account for background/foreground contamination by repeating the same procedure on sources from an equal-area region outside 3$r_{h}$ from the center and subtracting these values from the Gru~II-centered values. We obtain the best-fit distance modulus when the count of Gru~II sources is maximum, giving $m-M=18.54$ for \feh$=-2.2$ (143 sources), and $m-M=18.64$ for \feh$=-2.0$ (142 sources). 

Similarly, for Hor~II, we select all sources with $r$ $<=$ 24.5~mag within 1~$r_{h}$. We vary the distance modulus in steps of 0.025 mag from 18.5 to 20.5 mag, in line with the estimate by \citet{Kim2015a}, $m-M$ = 19.46$\pm$0.20~mag, and also account for background contamination. The best-fit distance moduli of Hor~II with different isochrones are: $m-M=19.27$ for age = 12.5~Gyr and \feh$=-2.2$ (54 sources), $m-M=19.29$ for age = 12.5~Gyr and \feh$=-2.0$ (53 sources), and $m-M=19.33$ for age = 13.5~Gyr and \feh$=-2.1$ (47 sources).

We quote the mean value from different isochrones as the final distance modulus for each satellite in Table~\ref{table:struct}. The statistical uncertainties on each fit are derived from a 100-iteration bootstrap resampling, and are added in quadrature with the standard deviation among the isochrone-based values to yield the final uncertainties reported in Table~\ref{table:struct}. As a check, we also estimated the distance moduli using the Dartmouth Stellar Evolution Database \citep[DSEP;][]{Dotter2008} isochrones, and the results are found to be consistent with the PARSEC-based estimates within 1$\sigma$ of the quoted uncertainties in Table~\ref{table:struct}.

\begin{table*}[]
\centering
\caption{Structural Properties of Gru~II \& Hor~II}
\begin{tabular}{l l l c}
 \hline \hline
 Parameter & Gru~II & Hor~II & Ref.\\
 \hline
 
  $\alpha_{\textrm{2000}}$ (deg) &
  \gruRA$^{+27\arcsec}_{-33\arcsec}$ &
  \horoRA$^{+20\arcsec}_{-19\arcsec}$ &
  0 \\

   $\delta_{\textrm{2000}}$ (deg) &
   \gruDEC$^{+24\arcsec}_{-20\arcsec}$ &
   \horoDEC$^{+15\arcsec}_{-14\arcsec}$ &
   0 \\

  $m-M$ (mag) &
  \gruDM$\pm$\gruDMerr &
  \horoDM$\pm$\horoDMerr &
  0 \\

   D (kpc) &
   \gruDKPC$\pm$\gruDKPCerr &
   \horoDKPC$^{+5.9}_{-5.5}$ &
   0 \\

   $M_{V}$ (mag) &
   \gruMV$\pm$\gruMVerr &
   \horoMV$\pm$\horoMVerr &
   0 \\

   $r_{h}$ (arcmin) &
   \gruRH$\pm$\gruRHerr &
   \horoRH$\pm$\horoRHerr &
   0 \\

   $r_{h}$ (pc) &
   \gruRHPHYS$\pm$\gruRHPHYSerr &
   \horoRHPHYS$^{+6}_{-5}$ &
   0 \\

   $\epsilon$ &
   \gruELLIP$\pm$\gruELLIPerr &
   \horoELLIP$^{+0.20}_{-0.16}$ &
   0 \\

   $\theta$ (deg) &
   \gruTHETA$^{+9}_{-12}$ &
   \horoTHETA$^{+44}_{-49}$ &
   0 \\

   \hline 

   $v_{sys}$ (km s$^{-1}$) &
   $-$110.0$\pm$0.5 &
   168.7$^{+12.9}_{-12.6}$ &
   1; 3 \\

   $\sigma_{v}$ (km s$^{-1}$) &
   $<$~2.0 &
   $<$~54.6 &
   1; 3 \\

   [Fe/H] (dex) &
   $-$2.51$\pm$0.11 &
   $-$1.87$^{+0.36}_{-0.50}$ &
   1; 3 \\

   $\sigma_{\text{[Fe/H]}}$ (dex) &
   $<$~0.45 &
   $<$~1.93 &
   1; 3 \\

   $\mu_{\alpha} \cos \delta$ (mas yr$^{-1}$) &
   \gruPMRA$\pm$\gruPMRAerr &
   \horoPMRA$\pm$\horoPMRAerr &
   2; 2 \\

   $\mu_{\delta}$ (mas yr$^{-1}$) &
   \gruPMDEC$\pm$\gruPMDECerr &
   \horoPMDEC$\pm$\horoPMDECerr &
   2; 2 \\

   $r_{peri}$ (kpc) &
   27.2$^{+8.4}_{-6.4}$ &
   68.2$^\dagger$ &
   2; 4\\

 \hline
\end{tabular}

\begin{tablenotes}
\small
\item $\alpha_{\textrm{2000}}$: Right Ascension (J2000.0). 
$\delta_{\textrm{2000}}$: Declination (J2000.0). 
$m-M$: distance modulus. 
$D$: distance of the galaxy in kpc. 
$M_{V}$: absolute V-band magnitude. 
$r_{h}$: elliptical half-light radius along the semi-major axis. 
$\epsilon$: ellipticity defined as $\epsilon=1-b/a$, where $b$ is the semi-minor axis and $a$ is the semi-major axis. 
$\theta$: position angle. 
$v_{sys}$: systemic radial velocity in the heliocentric frame. 
$\sigma_{v}$: velocity dispersion. 
[Fe/H]: mean metallicity. 
$\sigma_{\rm [Fe/H]}$: metallicity dispersion. 
$\mu_{\alpha}\cos\delta$: systemic proper motion in R.A. 
$\mu_{\delta}$: systemic proper motion in Dec. 
$r_{peri}$: orbital pericenter. 

$\dagger$ For Hor~II, the 16$^{th}$ percentile of the distribution of the orbital pericenter calculated using a MW potential perturbed by a heavy LMC is reported. It is also to be noted that the uncertainty on its total Galactocentric velocity exceeds 70~km~s$^{-1}$.

\item Reference codes: (0) This work; (1) \citet{Simon2020}; (2) \citet{Pace2022}; (3) \citet{Fritz2019}; (4) \citet{Battaglia2022}

\end{tablenotes}

\label{table:struct}
\end{table*}

\subsection{Structural Properties} \label{subsec:struct}

To determine the structural parameters of the dwarf satellites, we use a maximum likelihood technique to fit an exponential profile to the 2D distribution of stars \citep[][see also \citealt{Martin2008}]{Sand2009}. For each dwarf, stars which are brighter than r = 24.0~mag are selected if they are consistent with a PARSEC isochrone (12.5 Gyr, [Fe/H]=$-$2.2). This consistency is defined using a tubular selection around the isochrone, with a width derived from photometric uncertainties. To prevent the selection from becoming unrealistically narrow at bright magnitudes, a minimum uncertainty of 0.1 mag in $r$ is imposed. The resulting error-based envelope is then empirically scaled to match the observed CMD spread of each system: for Gru~II, a factor of 2.7 is adopted based on sources within 1~$r_{h}$, while for Hor~II, a factor of 2.5 is applied using sources within 2~$r_{h}$ to ensure adequate sampling. The exponential profile fit constrains the dwarf central position ($\alpha_{0}$, $\delta_{0}$), half-light radius ($r_h$), ellipticity ($\epsilon$), position angle ($\theta$) and the background surface density. Uncertainties on the parameters were constrained via 1000 bootstrap resampling iterations of the data. The algorithm accounts for regions where data is `missing', such as bad CCD chips and regions around saturated stars by estimating the effective observed area from random positions sampled across the survey footprint and retaining only those that fall within regions with usable imaging data. The structural parameters and associated uncertainties thus obtained are tabulated in Table~\ref{table:struct}.

\subsection{Absolute Magnitude}\label{subsec:abs_mag}

We determine the absolute magnitudes for both the galaxies by employing the procedure outlined in \citet{Martin2008} and subsequently used by \citet{Sand2009, Sand2010, Sand2012, Mutlu-Pakdil2018, Casey2025}. For each galaxy, we first create an artificial CMD consisting of $\sim$45,000 stars, populated using luminosity functions generated from the PARSEC database for a metallicity of $\text{[Fe/H]}=-2.2$ and assuming a \citet{Salpeter1955} initial mass function (IMF). The simulated CMD is designed to be consistent with the photometric uncertainties and completeness limits of our observational data. We then randomly sample $N$ stars from this artificial CMD, where $N$ is derived from the exponential profile fits (see Section \ref{subsec:struct}). This sampling was restricted to the same magnitude range used to derive the structural parameters. We estimate the total luminosity by summing the fluxes of these $N$ selected stars, along with the flux from the faint, unresolved stellar component derived based on the adopted luminosity function. This process is repeated 1,000 times. We adopt the mean value from these 1,000 realizations as the final absolute magnitude and the standard deviation as its statistical uncertainty. Additionally, we account for resolved HB stars by explicitly adding the fluxes of HB candidates identified within 3~$r_{h}$ of each satellite. For Gru~II, which has 11 HB candidates, we also empirically estimate the possible number of contaminants. We compute the number of HB candidates per unit area of active pixels outside 3~r$_{h}$ and multiply this value by the area of the 3~r$_{h}$ ellipse, yielding 1 contaminant. We account for the effect of this estimated contaminant by incorporating it into the uncertainty budget for the absolute magnitude of Gru~II. To calculate the total uncertainty, we propagate the errors associated with the distance modulus and the number of stars ($N$), also accounting for the contaminant HB flux contribution (for Gru~II). We do this by repeating the entire 1,000-realization process 100 additional times, allowing both the distance modulus and $N$ to vary according to their respective errors and by removing the estimated HB contaminant fraction each time. The final uncertainty on the absolute magnitude is obtained by adding these propagated errors in quadrature with the statistical uncertainty from the initial 1,000 realizations. This procedure yields absolute magnitudes of M$_V$ = \gruMV$\pm$\gruMVerr~mag for Gru~II and M$_V$ = \horoMV$\pm$\horoMVerr~mag for Hor~II, as listed in Table \ref{table:struct}. We note that the procedure adopted here assumes a single-component exponential profile for the UFDs \citep[see][ for a related discussion]{ Andersson2025}. 

\subsection{Search for Extended Structure} \label{subsec:ext_struct}

Taking cues from previous studies that have detected hints of tidal disruption in MW satellites \citep[e.g.,][]{Sand2009, Munoz2010}, we search for signs of extended structure in our targets. In this context, Gru~II is particularly interesting as \citet{Pace2022} predict that this satellite may be potentially disrupting.

\begin{figure*}[t]
    \centering

    \begin{minipage}[t]{0.49\textwidth}
        \centering
        \includegraphics[height=8.55cm, trim=35 18 28 60, clip]{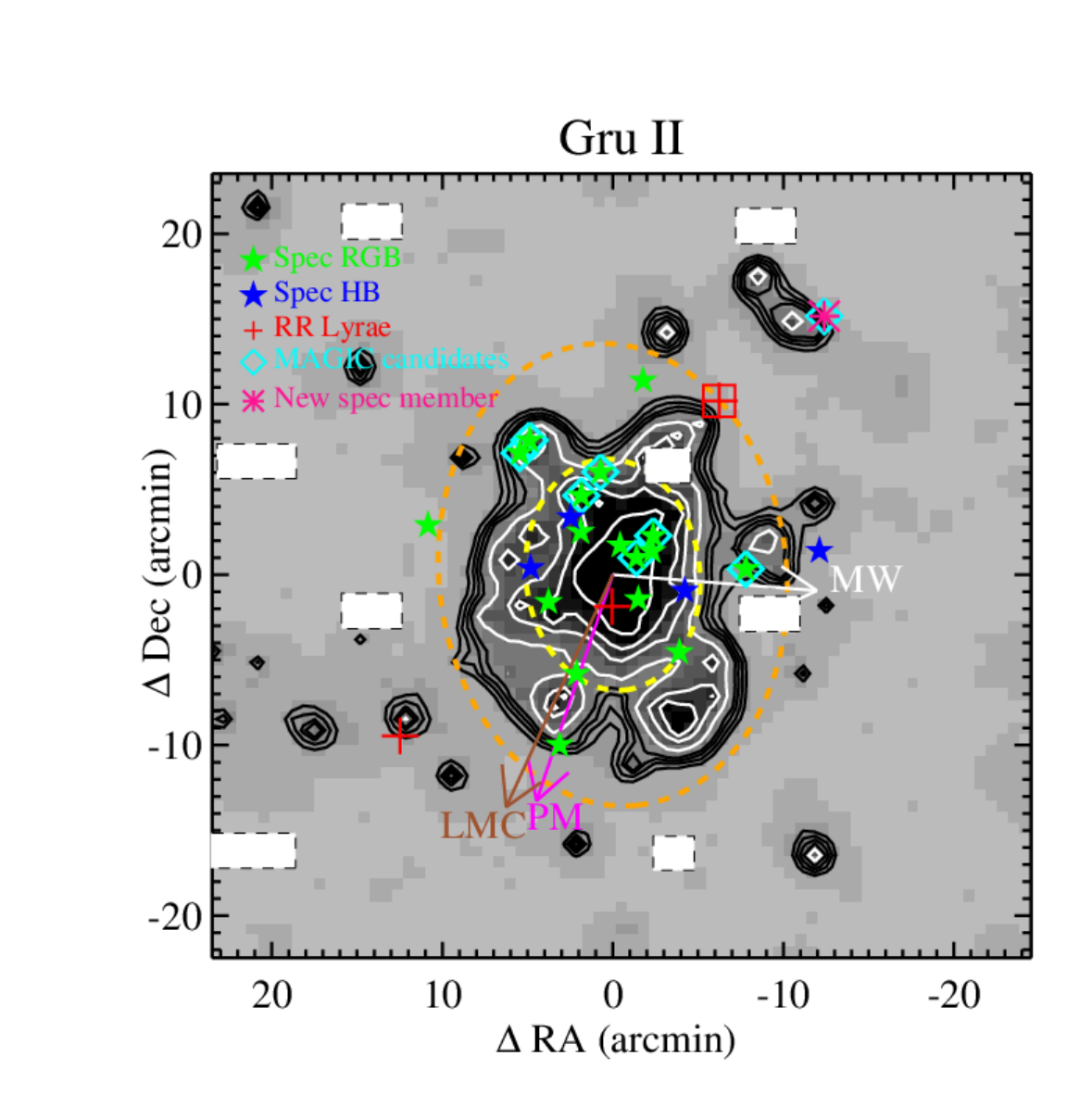}
        \vspace{2pt}
        (a) Matched-filter map
    \end{minipage}
    \hfill
    \begin{minipage}[t]{0.49\textwidth}
        \centering
        \includegraphics[height=8.08cm]{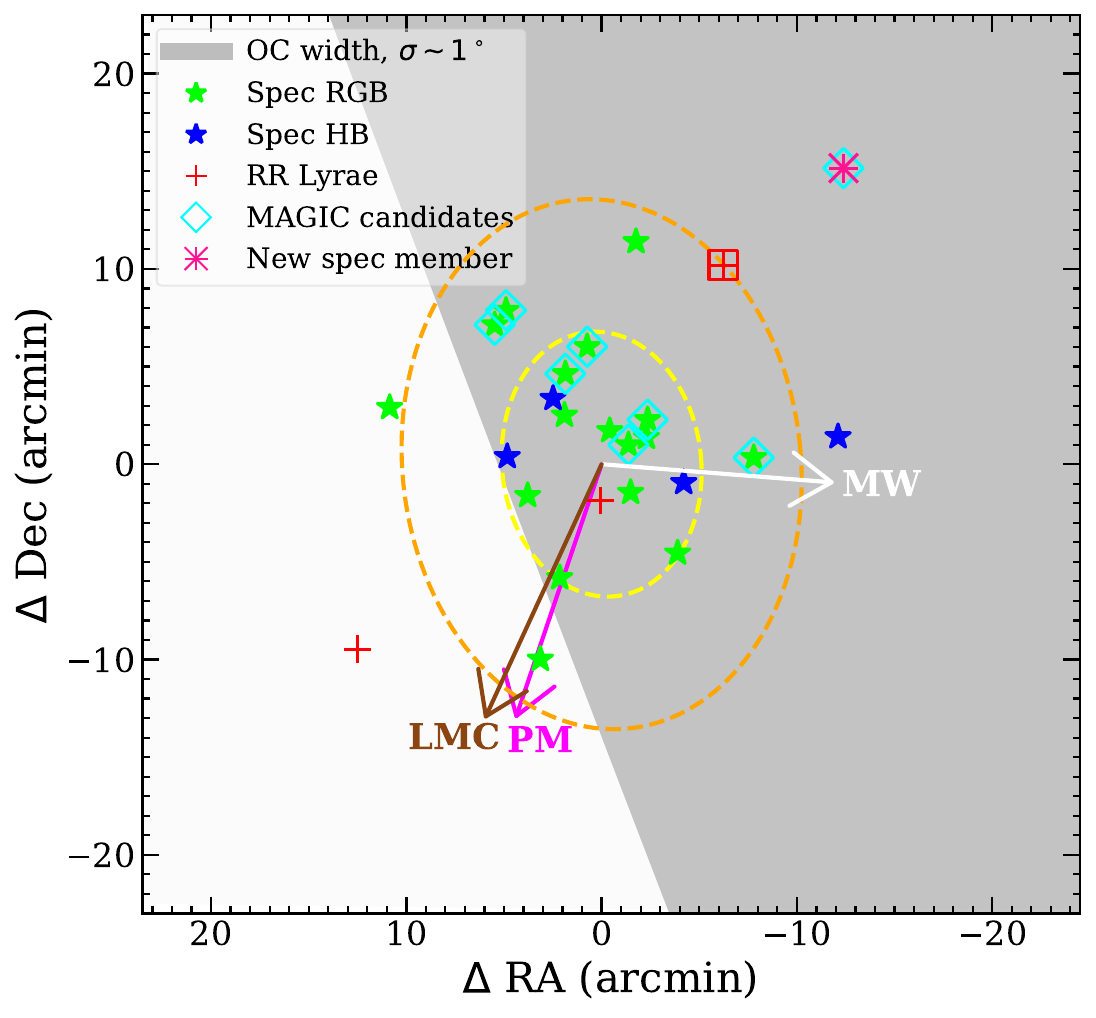}
        \vspace{2pt}
        (b) Gru~II--OC stream overlap
    \end{minipage}

    \caption{
Panel (a): The matched-filter map for Gru~II where the image center corresponds to the RA and DEC reported in Table~\ref{table:struct}. The contour levels above the root mean square of the background corresponding to $4\sigma$, $5\sigma$, $6\sigma$, $7\sigma$, $10\sigma$, $15\sigma$, $20\sigma$ and $30\sigma$ are overplotted. The white and brown arrows denote the direction to the Galactic center and LMC, respectively. The magenta arrow marks the direction of solar-reflex-corrected proper motion of Gru~II. The yellow and orange dashed ellipses correspond to 1 and 2~$r_h$, respectively. The RR Lyrae star of Gru~II and the two RR Lyrae stars of the OC stream from \citet{Martinez-Vazquez2019}, which fall in our FOV, are marked with a red cross enclosed with a square and a simple red cross, respectively. The spectroscopic members from \citet{Simon2020} are overlaid in green (RGB) and blue (HB). The MAGIC-selected candidates (CMD+proper motion+\feh$_{\text{MAGIC}}$-selected sources; see Section~\ref{subsec:spectra}) falling within the FOV are shown using cyan diamonds. The member identified in Section~\ref{subsec:spectra} through MagE spectroscopy is marked with a pink asterisk. The white patches in the image represent faulty CCDs. Panel (b): Plot showing the projected overlap between Gru~II and the OC stream whose approximate width ($\sigma$~=~1$^\circ$) is indicated by the gray shaded region. Symbols are as in panel (a).}
    \label{fig:ext_struct_Gru}
\end{figure*}

\begin{figure*}
    \centering
\includegraphics[width=0.75\textwidth, height=0.55\textwidth]{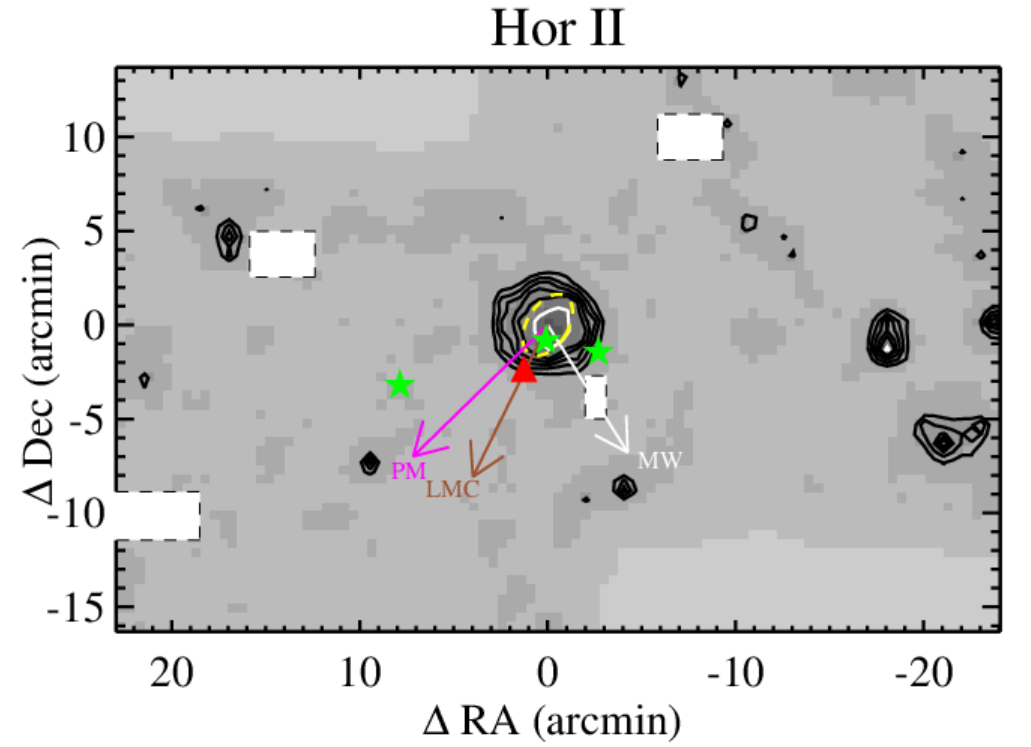}
\caption{The matched-filter map for Hor~II where the image center corresponds to the RA and DEC reported in Table \ref{table:struct}.  The contour levels above the root mean square of the background which corresponds to $3\sigma, 4\sigma$, $5\sigma$, $6\sigma$, $7\sigma$, $10\sigma$ and $15\sigma$. The yellow dashed ellipse corresponds to 1~$r_h$. The arrows are the same as in Figure~\ref{fig:ext_struct_Gru}. The likely spectroscopic members and candidate from \citet{Fritz2019} are overlaid in green and the new candidate member described in Sec.~\ref{subsec:gaia_members} is shown using a red triangle.}
    \label{fig:ext_struct_Horo}
\end{figure*}

We employ a matched-filter algorithm \citep{Rockosi2002}, a standard technique for investigating UFDs for signs of extensions \citep[e.g.,][]{Sand2012, Mutlu-Pakdil2018, Casey2025}. This method optimally enhances stars that are consistent with the expected CMD of each system, while suppressing foreground and background contaminants. Our signal CMD utilizes the simulated stars detailed in Section~\ref{subsec:abs_mag}, while the background CMD was derived from the observed sources located outside $3$~r$_{h}$ of the targets. Following \citet{Sand2012}, we adopt a simulated CMD rather than the observed dwarf galaxy CMD, since the latter can be sparsely populated and contaminated by foreground/background sources. The resulting matched-filter maps including all the pointings for both satellites are shown in Figure~\ref{fig:ext_struct_Gru} (a) and Figure~\ref{fig:ext_struct_Horo}. The input data was spatially binned to 40\arcsec\ pixels and 30\arcsec\ pixels respectively for Gru~II and Hor~II and then smoothed with a Gaussian width of 1.5 times the respective pixel size. IDL's MMM routine was used to estimate the background and variance of the smoothed maps. 

In both the matched-filter maps in Figures~\ref{fig:ext_struct_Gru}(a) and ~\ref{fig:ext_struct_Horo}, the main bodies (within 1~$r_h$) appear as clear over dense regions above the background in the maps. For context, the white arrows indicate the direction to the Galactic center, the brown arrows indicate the Large Magellanic Cloud (LMC) direction and the magenta arrows denote the direction of the solar-reflex-corrected proper motion vectors of the UFDs (indicating the orbital direction; \citealt{Pace2022}). For Gru~II, the spectroscopically confirmed member stars \citep{Simon2020} are overlaid in green (RGB) and blue (HB). Additionally, the MAGIC-selected candidates (CMD+proper motion+\feh$_{\text{MAGIC}}$-selected sources; see Section~\ref{subsec:spectra}) are overlaid in cyan. The new member identified in Section~\ref{subsec:spectra} through MagE spectroscopy is marked with a pink asterisk symbol. Finally, the Gru~II and OC stream RR Lyrae star members from \citet{Martinez-Vazquez2019} are marked with a red cross enclosed in a square and simple red crosses, respectively. For Hor~II, the likely spectroscopic members \citep{Fritz2019} are overlaid as green star symbols.

From Fig.\ref{fig:ext_struct_Gru}(a), we find that Gru~II exhibits an asymmetric and clumpy morphology. The highest-significance features (white contours; $\gtrsim$10$\sigma$ above the background) are not confined to the central body but instead trace multiple clumpy overdensities connected to it. These features persist across different smoothing scales (1.5$\arcmin$ and 2$\arcmin$) and for varying limiting magnitudes of the signal CMD used to construct the matched-filter map, supporting their robustness. Several spectroscopically-confirmed members are also observed to lie along or near these contours. We additionally detect a few isolated relatively compact overdensities (“nuggets”) outside the $2~r_h$ ellipse. However, visual inspection of these regions reveals concentrations of faint background galaxies, suggesting that these are unrelated to the UFD.

In contrast, the matched-filter map of Hor~II (Fig.~\ref{fig:ext_struct_Horo}) shows only low-significance extensions connected to the main body. While several localized overdensities are present in the map, these are spatially disconnected from the satellite. Visual inspection indicates that these apparent “nuggets” coincide with clusters of background galaxies, suggesting that they do not represent genuine stellar substructure associated with Hor~II. We discuss these maps and their implications further in Section~\ref{sec:discussion}.

\subsection{Potential Member Stars in Hor~II from Gaia} \label{subsec:gaia_members}

To facilitate future spectroscopic follow-up, we search for potential new members in Hor~II utilizing our deep Megacam photometry, precise astrometry from \textit{Gaia}~DR3 and the information on likely spectroscopic members in the literature. We follow \citet{Casey2025} and first cross-identify the likely spectroscopic members from \citet{Fritz2019} in our catalog. We then select sources from our CMD along the best-fit isochrone, as in Section~\ref{subsec:struct}, but limited to $r$ $<$ 21.5~mag and with the tubular boundary widened to a factor of 9 to encompass all likely spectroscopic members and candidates in the selection. We cross-match this subset to \textit{Gaia}~DR3 with a maximum match radius of 0.5\arcsec, retaining only those sources that satisfy the following criteria: {\fontfamily{cmtt}\selectfont ruwe}~$<1.4$, {\fontfamily{cmtt}\selectfont astrometric\_excess\_noise}~$<2$, and stars which are consistent with zero parallax ($\varpi - 3\sigma_{\varpi} < 0$). We further select sources that are consistent within 3$\sigma$ of the mean proper motion of the two likely spectroscopic members from \citet{Fritz2019} with \textit{Gaia}~DR3 proper motions. This yields one candidate in Hor~II, listed in Table~\ref{tab:horoII_gaia_mem} and shown using a red triangle in Fig.~\ref{fig:ext_struct_Horo}. We note that this star, located within $\sim$~2~r$_h$, has also been identified as a high-probability member ($\sim$97\%) in previous catalogs based on \textit{Gaia}~DR2 and EDR3 proper motions \citep{Pace2019, Pace2022, Battaglia2022}. 
We note that this source has no entry in the MAGIC catalog, which is why it did not appear in our search for low metallicity members in the outskirts of Hor~II (see Section~\ref{subsec:spectra}). 
The star was missed due to incompleteness from undithered MAGIC coverage of Hor~II (i.e., spatially coincident with a chip-gap);  consequently, we are unable to assess whether its photometric metallicity is consistent with Hor~II membership.

\begin{table*}
\caption{Details of the candidate member star identified in Hor~II.}\label{tab:horoII_gaia_mem}
\begin{minipage}[b]{0.95\linewidth}\centering
\begin{tabular}{lccccccc}
\tablewidth{0pt}
\hline
\hline
Gaia ID & R.A. & Dec & $g$ & $r$ & $\mu_{\alpha}\cos\delta$ & $\mu_{\delta}$ & Type\\
{} & (deg) & (deg) & (mag) & (mag) & (mas yr$^{-1}$) & (mas yr$^{-1}$) & \\
\hline
4749213280723458816 & 49.166621 & $-$50.046902 & 19.81 & 19.20 & 1.044 $\pm$ 0.210 & $-$0.640 $\pm$ 0.269 & RGB \\
\hline
\end{tabular}
\begin{tablenotes}
\small
\item Cols. (1)--(8) list the \textit{Gaia}~DR3 Source ID, Right Ascension, Declination, $g$-band and $r$-band magnitudes from our Megacam photometry, \textit{Gaia}~DR3 proper motions and the source type, respectively.

\end{tablenotes}
\end{minipage}
\end{table*}

\section{Discussion} \label{sec:discussion}

\begin{figure*}[t!]
    \centering
\includegraphics[width=0.8\textwidth, height=0.6\textwidth]{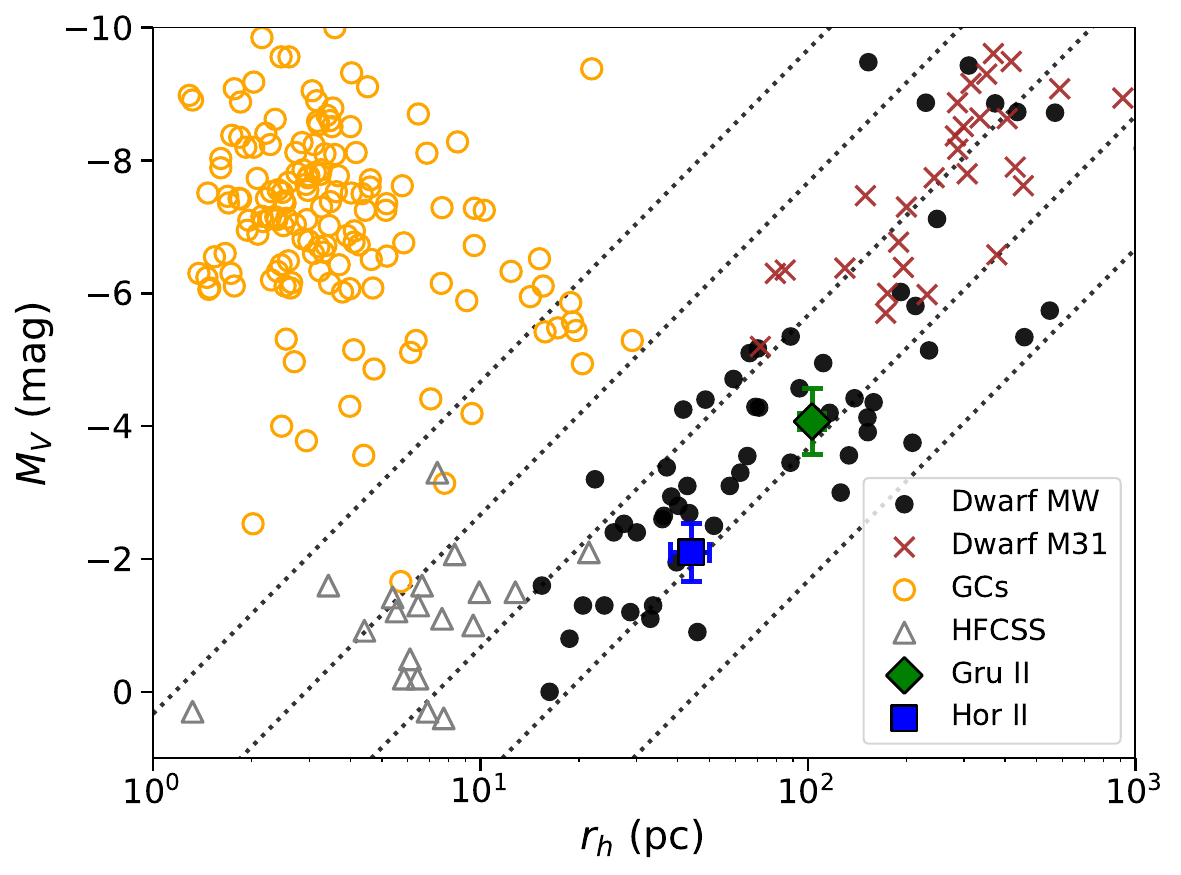}
\caption{A plot of the half-light radius ($r_{h}$) versus the absolute magnitude ($M_{V}$) for Local Group dwarfs and GCs. Gru~II and Hor~II are represented using a green diamond and blue square, respectively. The other MW satellites are shown with black circles and brown x's represent the M31 dwarf satellites. The orange open circles are MW GCs, and gray triangles are the HFCSSs. The dotted lines represent lines of constant surface brightness at $\mu$ = \{24.75, 26.75, 28.75, 30.75, 32.75\} mag~arcsec$^{-2}$. The data in this figure are adopted from the Local Volume Database compiled by \citet{Pace2025}.}
    \label{fig:size_lum}
\end{figure*}

In Fig.~\ref{fig:size_lum}, we compare the positions of Gru~II (green diamond) and Hor~II (blue square) in the size-luminosity ($r_{h}$ versus $M_{V}$) plane with respect to other MW satellites, globular clusters (GCs), M31 dwarfs and hyper-faint compact stellar systems (HFCSSs)\footnote{Also referred to as Ultra-Faint Compact Satellites \citep[UFCSs; ][]{Cerny2026}. }. The data for all stellar systems in this figure, except Gru~II and Hor~II, are taken from the Local Volume Database \citep[LVDB v1.1.0;][]{Pace2025} and the source references are listed in Appendix~\ref{sec:appendix}. We find that both Gru~II and Hor~II fall well within the parameter space occupied by the MW dwarf satellites. In the following subsections, we discuss each of our targets' derived properties in detail.

\subsection{Gru~II}

Our deep photometry of Gru~II in multiple pointings reveals a richly-populated main sequence (MS), a clear RGB and several HB candidates within 3~$r_{h}$ (left panel of Fig.~\ref{fig:cmds}~(a)). The satellite's stellar population is consistent with being old (age $\sim$ 12.5~Gyr), and very metal-poor\footnote{The isochrone adopted here has \feh = $-$2.2 (the most metal-poor PARSEC isochrone), while the spectroscopic study by \citet{Simon2020} find $\langle$\feh$\rangle$ = $-2.51\pm0.11$ for Gru~II.}. This is further supported by the \textit{HST}-based star formation history (SFH) analysis of \citet{Durbin2025}, who found that Gru~II experienced very early quenching, consistent with an ancient stellar population. Our distance estimate for this satellite (D = \gruDKPC ~$\pm$ \gruDKPCerr~kpc) is in good agreement with the estimate from the discovery paper by \citealt{DW2015}{ (D = 53 $\pm$ 5~kpc)}, and with the RR~Lyrae-based distance from \citealt{Martinez-Vazquez2019}{ (D = 55 $\pm$ 2~kpc)}. Out of the seven RR Lyrae reported in the field of Gru~II \citep{Martinez-Vazquez2019, Tau2024}, three fall within our FOV. But, as suggested by \citet{Martinez-Vazquez2019}, only the faintest one likely belongs to the UFD (see Fig.~\ref{fig:cmds} and Fig.~\ref{fig:ext_struct_Gru}). 
We also spectroscopically confirm the presence of one RGB member just outside 3\,r$_h$, northeast of the system, initially identified by metallicity-sensitive MAGIC photometry, suggestive of an associated stellar population existing beyond the clumps connected to the main body of the system in the matched filter map. To quantify the consistency of this star with Gru~II, we compute the membership score from \citet{Tolstoy2023}. The membership score, $z_4$, is effectively a 3\,$\sigma$ selection in velocity, proper motion, and parallax space and accounts for correlations between parameters. We find $z_4=3.8$, consistent with Gru~II as members will have $z_4<16.3$.

Using maximum likelihood analysis, we obtain robust measurements of the structural parameters of Gru~II (see Table~\ref{table:struct}). Our measurements are consistent with the discovery estimates of \citet{DW2015} within 1\,$\sigma$. We also find good agreement with the re-analysis of the discovery data by \citet{Simon2020}, with all directly comparable parameters agreeing within the joint 1\,$\sigma$ uncertainties except for r$_h$ (agrees within $\sim$~1.3\,$\sigma$), while yielding improved precision for some of the parameters. Our values of the $r_{h}$ and $\epsilon$, are consistent with those reported by \citet{MW2020} within the joint 1\,$\sigma$ uncertainties, despite the use of different models (exponential versus Plummer). The position angle derived in this work lies well within the large uncertainty range of their estimate. We derive an absolute V magnitude of $M_{V}$ = $-$4.07 $\pm$ 0.50, which is in good agreement with the values reported by \citealt{DW2015} { ($M_{V}$ = $-$3.90 $\pm$ 0.22)} and \citealt{Simon2020}{ ($M_{V}$ = $-$3.50 $\pm$ 0.3)}. With our measured values, Gru~II lies among the population of other confirmed MW dwarf satellites in the size–luminosity plane (see Fig.\ref{fig:size_lum}), lending support to its dwarf galaxy interpretation \citep{DW2015, Simon2020}.

The matched-filter map of Gru~II (see Fig.\ref{fig:ext_struct_Gru}(a)) reveals the prominent central body of the system together with an asymmetric morphology characterized by multiple clumpy overdensities connected to the main body. These connected features are detected at significance levels exceeding 10\,$\sigma$ above the background and persist across different smoothing scales and limiting magnitudes used to construct the signal CMD for the matched-filter map, supporting their robustness. Many of the spectroscopically confirmed members and the MAGIC-selected candidates (see Section~\ref{subsec:spectra}) are observed to lie along or near these contours, providing further support that these overdensities are associated with Gru~II. We note that all of the isolated nuggets elsewhere in the map spatially coincide with concentrations of background galaxies and are therefore not associated with Gru~II.

The morphology revealed from our deep imaging provides important context for interpreting the dynamical state of Gru~II. The system is unusual in exhibiting a very small velocity dispersion upper limit \citep[$\sigma_v <2.0~\mathrm{km~s^{-1}}$ at 95.5\% confidence;][]{Simon2020}. Despite this, the physical size and luminosity of Gru~II (see Table~\ref{table:struct} and Fig.\ref{fig:size_lum}) are consistent with those of known MW dwarf satellite galaxies. The asymmetric morphology and clumpy features connected to the main body identified in our matched-filter map suggest that Gru~II may have experienced external perturbations. If the system is affected by tidal processes, then the assumption of dynamical equilibrium may not hold, implying that dynamical mass estimates inferred from the velocity dispersion should be interpreted cautiously \citep{Kravtsov2004, Penarrubia2008, Kupper2017}. 

The orbital and environmental context of Gru~II further supports the possibility that the system has experienced tidal perturbation. \citet{Pace2022} identified Gru~II as a system potentially undergoing tidal disruption based on its relatively low average density with respect to the average MW density at the pericenter of the UFD. Gru~II also has a relatively small inferred pericentric distance (24--31~kpc) and low central density \citep{Pace2022, Battaglia2022}. The system may have interacted with or been captured by the LMC within the past $\sim200$ Myr \citep{Erkal2020, Battaglia2022, CorreaMagnus2022}. This interpretation is further supported by \citet{Santos-santos2021}, who identified Gru~II as a likely LMC associate. Together with the asymmetric features revealed by our deep imaging, these orbital properties suggest that tidal effects may have contributed to shaping the observed stellar distribution of Gru~II. The complex morphology of the matched-filter map, including structures that are not confined to a single preferred direction, may additionally reflect a combination of projection effects and multiple dynamical influences in a tidally evolving system \citep[e.g.,][]{Sato2025, Yang2026}.

Gru~II also overlaps with the OC stream spatially \citep{Koposov2019, Koposov2023} and shares a similar proper motion \citep{Shipp2019}. Panel (b) of Fig.~\ref{fig:ext_struct_Gru} illustrates this projected overlap within our imaging field with the gray shaded region\footnote{from the PYTHON package GALSTREAMS \citep{Mateu2023}}  representing the approximate width of the stream \citep[$\sigma$~=~1$^\circ$; ][]{Koposov2023}, although the center of the stream track itself lies outside our FOV. Despite this projected and kinematic overlap, other properties argue against a direct association. The radial velocity of Gru~II and the predicted velocity of the OC stream differ by $\sim90~\mathrm{km~s^{-1}}$ \citep{Simon2020}, while \citet{Martinez-Vazquez2019} found Gru~II to be located $\sim$10~kpc farther away than the stream. Although overlap in the proper motion space complicates member identification, the OC population is naturally incorporated into the background used in our matched-filter analysis. Furthermore, the signal CMD adopted for the matched-filter map is optimized for the old, very metal-poor stellar population of Gru~II at its measured distance (D = \gruDKPC ~$\pm$ \gruDKPCerr~kpc), whereas the OC stream is relatively more metal-rich \citep[\feh$_{\text{mean}}$ = $-$1.9; ][]{Koposov2023, Hawkins2023} and located $\sim$10~kpc closer \citep{Martinez-Vazquez2019, Koposov2023}. Taken together, these results strongly suggest that the clumpy features connected to the main body identified in Fig.~\ref{fig:ext_struct_Gru}(a) are predominantly associated with Gru~II rather than arising from OC stream contamination.

Higher ellipticities in UFDs have often been associated with tidal interactions \citep[e.g.,][]{Munoz2010, Kupper2017}, although a number of systems such as Tucana~III and Crater~II with clear tidal signatures show only modest ellipticities \citep[e.g.,][]{Munoz2008, DW2015, Torrealba2016a, Mutlu-Pakdil2018}. The measured ellipticity of Gru~II ($\epsilon$ = \gruELLIP$\pm$\gruELLIPerr) therefore does not argue against the possibility that the system has experienced tidal effects. At the same time, tidal stripping is not the only possible explanation for the asymmetric and clumpy morphology identified in our matched-filter map. Such morphological complexity in UFDs has additionally been attributed to early mergers, bursty feedback, or captured field stars \citep[e.g.,][]{Chiti2021b, Tarumi2021, Penarrubia2024}. Nevertheless, the combination of previous orbital studies and the asymmetric, clumpy morphology identified here supports the possibility that tidal effects have influenced the evolution of Gru~II. Our deep, wide-field imaging thus reveals structural complexity that was not previously accessible, providing new insight into Gru~II.

Beyond the clumpy morphology identified in the matched-filter map, we also find evidence that Gru~II contains member stars at large projected radii. Using a combination of \textit{Gaia}~DR3 proper motions, CMD selection, and MAGIC photometric metallicities (Fig.\ref{fig:cahkselection_GruII}), we identified a candidate member just outside $\sim$~3r$_{h}$, which we subsequently confirmed spectroscopically with MagE. This result demonstrates that bona fide Gru~II stars are present at least to projected radii of $\sim$~3r$_{h}$. Previous investigations of the outskirts of Gru~II have also been carried out using \textit{Gaia}~DR3 astrometry and photometry. For example, \citet{Jensen2024} applied a new algorithm to identify outer stellar components in MW satellites and reported evidence for an additional stellar component around Gru~II. The inclusion of photometric metallicity information in the MAGIC selection provides additional leverage for removing contaminants that would otherwise appear consistent with Gru~II based on proper motions alone (e.g., see right panel of Fig.~\ref{fig:cahkselection_GruII}). At present, however, the MAGIC coverage around Gru~II is spatially incomplete, preventing a direct comparison of the outer stellar distribution. Deeper imaging,  more complete photometric metallicity coverage and ultimately further spectroscopy are needed to make conclusive statements about the extended density profile of Gru~II.

\subsection{Hor~II}

The CMD of Hor~II shows a clear MS truncated at the turn-off and a sparse RGB (left panel of Fig.~\ref{fig:cmds}~(b)). We find two HB candidates within 3~r$_{h}$. Similar to Gru~II, Hor~II's  stellar population is old and metal-poor, with \textit{HST}-based SFH measurements by \citet{Durbin2025} likewise indicating very early quenching. Hor~II does not currently have any confirmed RR~Lyrae stars, consistent with its sparsely populated HB, suggesting that there are likely none within the main body. This is also broadly consistent with the empirical $N_{\rm RRL}$ versus $M_V$ relation presented by \citet{Martinez-Vazquez2019,Martinez-Vazquez2023}, which shows that UFDs fainter than $M_V = -3$ typically host $\lesssim 1$ RR~Lyrae, with several systems having none detected at all \citep[e.g., Draco~II, Carina~III]{Vivas2020, Tau2024}. Given the low luminosity of Hor~II ($M_V$ = \horoMV$\pm$\horoMVerr), the absence of confirmed RR~Lyrae stars is therefore not unexpected. \citet{Vivas2020}, using \textit{Gaia}~DR2 data, and more recently \citet{Tau2024}, using \textit{Gaia}~DR3, likewise found no RR~Lyrae stars associated with Hor~II. Since the expected \textit{Gaia} $G$-band magnitude of the HB of Hor~II is 19.93~mag \citep{Tau2024}, well above the \textit{Gaia} limiting magnitude of $G \sim 21$~mag, this further supports the conclusion that Hor~II likely lacks RR~Lyrae stars in the main body.

Our structural parameter measurements for Hor~II (Table~\ref{table:struct}) are consistent with the discovery values reported by \citet{Kim2015a} at the 1\,$\sigma$ level. They are also in good agreement with the more recent measurements of \citet{Richstein2024} based on HST imaging, with the half-light radius agreeing at the $\sim$1.8\,$\sigma$ level and all other parameters consistent within 1\,$\sigma$.
We find an absolute magnitude of $M_V$ = \horoMV$\pm$\horoMVerr, which agrees with the discovery value of $M_{V} = -2.6^{+0.2}_{-0.3}$ \citep{Kim2015a} within the joint 1\,$\sigma$ uncertainties, and is in excellent agreement with the value derived by \citet{Richstein2024}, $M_{V} = -2.1 \pm 0.2$. Our distance estimate of \horoDKPC$^{+5.9}_{-5.5}$~kpc is consistent within 1\,$\sigma$ with the value reported by \citet{Kim2015a}, $D = 78^{+8}_{-7}$ kpc. With these parameters, Hor~II occupies a region of the size–luminosity plane consistent with other ultra-faint MW satellites (Fig.~\ref{fig:size_lum}), supporting its tentative classification as a dwarf galaxy \citep{Kim2015a, Baumgardt2022}.

The dynamical properties of Hor~II remain poorly constrained because its member sample is extremely sparse and still uncertain. Spectroscopic follow-up by \citet{Fritz2019} identified only three candidate member stars, which continue to form the basis of current line-of-sight velocity estimates. Using a contamination model based on foreground expectations and the velocity distribution, \citet{Fritz2019} classified two of these stars as very likely members (membership probability = 1.0) while the third star remains a candidate member based on a probability membership of 0.5. The limited number of stars and their relatively poor velocity precision leave the systemic velocity of Hor~II among the most uncertain in their sample. \citet{Pace2019} subsequently examined this system using \textit{Gaia}~DR2 astrometry and identified five candidate members, of which three were assigned higher membership probabilities. Although these candidates cluster in proper-motion space, they noted that the uncertainties were too large to robustly establish them as belonging to a common system and that the satellite posterior remained poorly constrained. They further argued that the limited spectroscopic sample of \citet{Fritz2019} did not provide sufficiently strong constraints to robustly establish the heliocentric velocity of Hor~II. Subsequent \textit{Gaia}~EDR3-based analyses found that the Hor~II signal in proper-motion space became more significant while largely recovering the same candidate members \citep{Pace2022}. However, the small number of stars with \textit{Gaia} detections continued to limit robust constraints on the system kinematics and orbit. \citet{Battaglia2022} likewise noted that Hor~II remains one of the most challenging systems in their sample, with the systemic line-of-sight velocity still among the most uncertain and individual membership probabilities remaining sensitive to the inclusion of spectroscopic information. Consequently, a possible association of Hor~II with the LMC remains tentative.

In the context above, our deep photometry combined with \textit{Gaia}~DR3 astrometry allows us to revisit the member selection. Using CMD-based filtering and a 3\,$\sigma$ proper motion selection around the known members, we recover one high-probability candidate (see Table~\ref{tab:horoII_gaia_mem} and Fig.~\ref{fig:ext_struct_Horo}) previously identified by \citet{Battaglia2022} with membership probability $\sim$0.97, providing an independent consistency check of the current member sample. Our current narrow-band CaHK imaging of Hor~II, applied in a similar manner as for Gru~II, returns no proper motion-consistent candidate members that are in the very metal-poor regime (see Figure~\ref{fig:cahkselection_HorII}). This is due to incompleteness from undithered narrow-band photometric coverage; for instance, our high-probability candidate above unfortunately contains no record in the MAGIC catalog due to its location in a chip-gap. Moreover, we find that one of the candidate members in \citet{Fritz2019} fails our metallicity selection, due to a relatively high but uncertain photometric metallicity. Additional narrow-band CaHK coverage would provide an important independent complement to assess membership for the one high-probability candidate presented above, and for studies of Hor~II stars more generally. 

Our matched-filter analysis for Hor~II reveals only low-significance extensions, with apparent overdensities attributable to background galaxy contamination, indicating no clear evidence for coherent tidal substructure in Hor~II.

\section{Summary and Conclusions}\label{sec:summary}

We present deep, wide-field Magellan/Megacam imaging of the ultra-faint MW satellites Gru~II and Hor~II. Our imaging covers a FOV extending to $\sim$5 times the $r_{h}$ of Gru~II and $>$10 times that of Hor~II, with photometry reaching $\sim$3 magnitudes deeper than the discovery data from DES. These data enable more precise estimates of their distances, luminosities, and structural properties, while also allowing us to probe their faint outskirts with much greater sensitivity. Both systems show CMDs consistent with old, metal-poor stellar populations, as expected for UFDs, and occupy regions of the size–luminosity plane typical of MW satellites.

For Gru~II, our structural parameter measurements are broadly consistent with previous work, further supporting its classification as a dwarf galaxy. The most notable result of our analysis is the detection of asymmetric clumpy stellar morphology surrounding the main body, with multiple high-significance structures detected at levels exceeding $10\sigma$ above the background. These features persist across different smoothing scales and matched-filter selections, and are traced by both spectroscopically confirmed members and candidate members selected from photometric metallicities, supporting their association with the system. We also identify and spectroscopically confirm a distant member star located just outside $\sim3~r_h$, initially flagged by its low photometric metallicity from MAGIC, providing independent evidence for Gru~II stars at large projected radii.

The combination of the asymmetric morphology revealed by our deep imaging, the unusually low velocity dispersion reported by \citet{Simon2020}, and the orbital properties of Gru~II suggests that the system may not be fully dynamically relaxed. Previous studies have identified Gru~II as potentially susceptible to tidal effects and as a likely former associate of the LMC, consistent with a scenario in which environmental interactions have influenced its evolution. While the complex morphology observed in the matched-filter map does not uniquely require a tidal origin, the combination of structural, dynamical, and orbital evidence suggests that tidal interactions may have contributed to shaping the present-day stellar distribution of Gru~II. Our deep, wide-field imaging therefore provides new evidence that Gru~II possesses a more complex structure than previously recognized.

Hor~II appears comparatively regular. Its structural and photometric properties are consistent with earlier measurements, and we do not detect statistically significant extended stellar features at the depth of our data. Within the current sensitivity limits, there is no clear evidence for extended structure in this system. To support future spectroscopic studies, we recover one high-probability candidate member using a combination of photometric selection, \textit{Gaia} DR3 astrometry, and proper motion consistency with known members, consistent with previous identifications in the literature.

Overall, our analysis demonstrates that deep, wide-field imaging provides critical leverage in characterizing the faint outskirts of UFDs, revealing structural features that are not accessible in shallower data and offering new constraints on their dynamical states and evolutionary histories.

\section*{Acknowledgments}

%We thank the anonymous reviewer whose comments improved the content of the paper. 
D.J.S. and the Arizona team acknowledges support from NSF grant AST-2508746. 
A.C. is supported by a Brinson Prize Fellowship grant through the Brinson Foundation. B.M.P. acknowledges support from NSF grant AST-2508745. D.C. acknowledges support from NSF grant AST-2508747. W.C. gratefully acknowledges support from a Gruber Science Fellowship at Yale University. This material is based upon work supported by the National Science Foundation Graduate Research Fellowship Program under Grant No. DGE2139841. Any opinions, findings, and conclusions or recommendations expressed in this material are those of the author(s) and do not necessarily reflect the views of the National Science Foundation. The works of Y. Choi and P. Massana are supported by NOIRLab, which is managed by AURA under a cooperative agreement with the U.S. National Science Foundation.

This paper includes data gathered with the 6.5-meter Magellan Telescope located at Las Campanas Observatory, Chile. 

The DECam Local Volume Exploration Survey (DELVE; NOAO Proposal ID 2019A-0305, PI: Drlica-Wagner) is partially supported by Fermilab LDRD project L2019-011 and the NASA Fermi Guest Investigator Program Cycle 9 No. 91201.
This project used data obtained with the Dark Energy Camera (DECam), which was constructed by the Dark Energy Survey (DES) collaboration. Funding for the DES Projects has been provided by the U.S. Department of Energy, the U.S. National Science Foundation, the Ministry of Science and Education of Spain, the Science and Technology Facilities Council of the United Kingdom, the Higher Education Funding Council for England, the National Center for Supercomputing Applications at the University of Illinois at Urbana–Champaign, the Kavli Institute of Cosmological Physics at the University of Chicago, the Center for Cosmology and Astro-Particle Physics at the Ohio State University, the Mitchell Institute for Fundamental Physics and Astronomy at Texas A\&M University, Financiadora de Estudos e Projetos, Fundação Carlos Chagas Filho de Amparo à Pesquisa do Estado do Rio de Janeiro, Conselho Nacional de Desenvolvimento Científico e Tecnológico and the Ministério da Ciência, Tecnologia e Inovação, the Deutsche Forschungsgemeinschaft and the Collaborating Institutions in the Dark Energy Survey.
The Collaborating Institutions are Argonne National Laboratory, the University of California at Santa Cruz, the University of Cambridge, Centro de Investigaciones Enérgeticas, Medioambientales y Tecnológicas–Madrid, the University of Chicago, University College London, the DES-Brazil Consortium, the University of Edinburgh, the Eidgenössische Technische Hochschule (ETH) Zürich, Fermi National Accelerator Laboratory, the University of Illinois at Urbana-Champaign, the Institut de Ciències de l'Espai (IEEC/CSIC), the Institut de Física d'Altes Energies, Lawrence Berkeley National Laboratory, the Ludwig-Maximilians Universität München and the associated Excellence Cluster Universe, the University of Michigan, the National Optical Astronomy Observatory, the University of Nottingham, the Ohio State University, the OzDES Membership Consortium, the University of Pennsylvania, the University of Portsmouth, SLAC National Accelerator Laboratory, Stanford University, the University of Sussex, and Texas A\&M University.
Based in part on observations at Cerro Tololo Inter-American Observatory, National Optical Astronomy Observatory, which is operated by the Association of Universities for Research in Astronomy (AURA) under a cooperative agreement with the National Science Foundation.
Database access and other data services are hosted by the Astro Data Lab at the Community Science and Data Center (CSDC) of the National Science Foundation's National Optical Infrared Astronomy Research Laboratory, operated by the Association of Universities for Research in Astronomy (AURA) under a cooperative agreement with the National Science Foundation.

This work has made use of data from the European Space Agency (ESA) mission
{\it Gaia} (\url{https://www.cosmos.esa.int/gaia}), processed by the {\it Gaia}
Data Processing and Analysis Consortium (DPAC,
\url{https://www.cosmos.esa.int/web/gaia/dpac/consortium}). Funding for the DPAC
has been provided by national institutions, in particular the institutions
participating in the {\it Gaia} Multilateral Agreement.

\hfil 

\facilities{Magellan: Clay (Megacam), Magellan: Baade (MagE spectrograph), CTIO: Blanco (DECam), \textit{Gaia}}

\textit{Software:} IDL astronomy users library \citep{Landsman1993}, SExtractor \citep{sextractor1996}, numpy \citep{numpy2020}, pandas \citep{pandas2021}, astropy \citep{astropy2022}, Topcat \citep{Taylor2005}.

\appendix

\section{References for the LVDB Comparison Sample}\label{sec:appendix}

MW dwarf satellites - \citet{McConnachie2012, DW2015, Kim2015a, Koposov2015a, Martin2015, Crnojevic2016, DW2016, Torrealba2016a, Torrealba2016b, Carlin2017, Choi2018, Homma2018, Koposov2018, Munoz2018, Mutlu-Pakdil2018, Torrealba2018, Homma2019, Wang2019, Mau2020, MW2020, Simon2020, Cantu2021, Cerny2021, Ji2021, Richstein2022, Cerny2023a, Cerny2023b, Homma2023, Smith2023, Casey2025}

M31 satellites - \citet{Martin2016, McConnachie2006, Richardson2011, 	Higgs2021, Slater2015, Arias2025, Rhode2023, McConnachie2012, Collins2022, Smith2025, Ogami2024, Collins2024}

MW GCs - \citet{Baumgardt2018, Harris1996, Munoz2018, Simpson2019, Kobulnicky2005, Kurtev2008, Deras2023, Pallanca2023, Leanza2024, Loriga2025}

HCFSSs - \citet{Popova2025, Balbinot2013, Mau2020, Cerny2021, Cerny2023a, Cerny2023b, Cerny2023c, Tan2026, Conn2018, Luque2018, Torrealba2019, Homma2019, Kim2015a, Kim2015b, Munoz2018, Martin2016}

\end{document}